\documentstyle[emlines]{article}
\def\appendix#1{
\addtocounter{section}{1}
\setcounter{equation}{0}
\renewcommand{\thesection}{\Alph{section}}
\section*{Appendix \thesection\protect\indent #1}
\addcontentsline{toc}{section}{Appendix \thesection\ \ \ #1}
}

\newcommand{\0}{\frac{1}{x}}
\newcommand{\1}{\frac{1}{x-1}}

\newcommand{\5}{x}
\newcommand{\6}{(x-1)}
\newcommand{\7}{(x+\frac{n_0}{N-n_0})}
\newcommand{\8}{(x-\frac{N-n_0-n_\infty}{N-n_0})}
\newcommand{\9}{(x-\frac{n_0}{n_0-n_{\infty}})}
\newcommand{\A}{\dot{A}}

\def\a{\dot{a}}
\def\b{\dot{b}}
\newcommand{\p}{{\bf p}}
\newcommand{\k}{{\bf k}}
\newcommand{\r}{\rangle}
\def\l{\langle}

\newcommand{\q}{{\bf q}}

\newcommand{\R}{{\bf R}}
\def\D{\Delta}
\def\al{\alpha}
\def\c{\dot{c}}
\def\t{\tau}
\def\T{\Theta}
\def\z{\zeta}
\def\S{\Sigma}
\def\CS{{\cal S}}
\def\e{{\,\rm e}\,}
\def\eb{{\bf e}}

\def\be{\begin{equation}}
\def\la{\label}
\def\ee{\end{equation}}
\def\bea{\begin{eqnarray}}
\def\eea{\end{eqnarray}}

\def\d{\delta}
\def\g{\gamma}
\def\s{\sigma}
\newcommand{\cH}{{\cal H }}
\newcommand{\cL}{{\cal L }}
\newcommand{\Z}{{\bf Z}}
\begin{document}
\title{
Four graviton scattering amplitude from $S^N\large{{\bf R}}^{8}$
supersymmetric orbifold sigma model}
\author{G.E.Arutyunov\thanks{arut@genesis.mi.ras.ru}
\mbox{} and S.A.Frolov\thanks{frolov@genesis.mi.ras.ru}
\mbox{} \\
\vspace{0.4cm}
Steklov Mathematical Institute,
\vspace{-0.5cm} \mbox{} \\
Gubkin str.8, GSP-1, 117966, Moscow, Russia;
\vspace{0.5cm} \mbox{} 
}
\date {} 
\maketitle 
\begin{abstract}
In the IR limit the Matrix string theory is expected to be described 
by  the $S^N\R^{8}$ supersymmetric orbifold sigma model. 
Recently  Dijkgraaf, Verlinde and Verlinde proposed a vertex that may
describe the type IIA string interaction.   
In this paper using this interaction vertex we derive the
four graviton scattering amplitude from the orbifold model 
in the large $N$ limit. 
\end{abstract}

\section{Introduction}
According to the Matrix theory conjecture \cite{BFSS} the quantum 
mechanics of $N$ D-particles \cite{W1} of type IIA string theory 
in the large $N$ limit describes the eleven-dimensional dynamics
of M-theory \cite{W2,Sch}. In particular, some subspace of the Hilbert space
of the quantum mechanics model admits an interpretation in terms of 
the second-quantized Fock space of the M-theory states, and the 
S-matrix of the model is directly related to the scattering 
amplitudes of the M-theory particles. The consistency of this
conjecture was already examined in many ways (for recent review see \cite{B}).

Compactifying  Matrix theory on a circle one arrives at the
${\cal N}=8$ two-dimensional supersymmetric $SU(N)$ Yang-Mills model \cite{T}.
It was recently argued in \cite{M,BS,DVV} that in the large $N$ limit the
Yang-Mills theory describes non-perturbative dynamics of type IIA string
theory, and the Yang-Mills and string coupling constants were shown to 
be inverse to each other. The argumentation was based on the 
observation that in the IR limit the gauge theory is 
strongly coupled and the IR fixed point may be described by the 
${\cal N}=8$ supersymmetric conformal field theory on the orbifold 
target space $S^N\R^8$. In particular, it is known \cite{DMVV}
that the Hilbert space of the orbifold model
coincides (to be precise, contains) in the 
large $N$ limit with the Fock space of the free second-quantized type 
IIA string theory. 

Basing on the string interpretation of the Hilbert space of the 
orbifold model, Dijkgraaf, Verlinde and Verlinde (DVV) \cite{DVV} 
suggested that perturbative string 
dynamics in the first order in the string coupling constant can be 
described by the $S^N\R^{8}$ supersymmetric orbifold conformal model 
perturbed by an irrelevant operator of conformal dimension 
$(3/2,3/2)$. Moreover, they determined an explicit form of this operator 
and showed that it preserved the space-time supersymmetry
and nicely fitted the conventional formalism of the 
light-cone string theory.

The described sigma-model approach to the perturbative 
second-quantized string theory is not limited only to the type IIA 
strings. On the same grounds one may easily define the DVV 
interaction vertices for the sigma-model description of bosonic, 
heterotic \cite{R}, and type IIB strings.  

An important problem posed by the above-described stringy
interpretation of the $S^N$ orbifold sigma models is to obtain the
usual string scattering amplitudes directly from the models. 
The positive result would obviously provide a strong evidence 
that Yang-Mills models indeed capture  nonperturbative
string dynamics. This problem seems
to be nontrivial due to the nonabelian nature of the $S^N$ orbifold models.
In our previous paper \cite{AF} we obtained the four tachyon scattering
amplitude from the $S^{N}\R^{24}$ orbifold conformal field theory
perturbed by the bosonic analog of the DVV interaction vertex.

The aim of the present paper is to derive the four graviton scattering
amplitude for type II strings from the $S^{N}\R^{8}$ 
supersymmetric orbifold sigma model, the closed string interaction 
being described by the DVV interaction vertex. 
We treat in detail only the more complicated case of type IIA strings
since for type IIB strings the left- and right-moving sectors
are identical.

Our consideration of the scattering amplitudes
starts with defining incoming and outgoing asymptotic states
$|i\r$ and $|f\r$ that should be identified with
some states in the Hilbert space of the orbifold conformal field
theory. In CFT any state is created by some conformal field and,
therefore, the first step consists in finding the conformal fields
of the orbifold CFT corresponding to the asymptotic states.
Recall that the Hilbert space of the orbifold sigma model
is decomposed into the direct sum of Hilbert spaces of twisted 
sectors. Each twisted sector describes asymptotic states of 
several strings. The vacuum state of a twisted sector corresponds to
a ground state twist operator. If the orbifold sigma model 
originates from the IR limit of the Yang-Mills theory, then
the energy of all vacuum states should be the same and, therefore,
the conformal dimensions of the ground state twist operators 
must be equal. We will show that this is indeed the case
for the supersymmetric orbifold sigma model. In contrast,
in the bosonic case the conformal dimensions 
were found \cite{AF} to be different and, therefore, the bosonic 
sigma model does not describe the IR limit of the Yang-Mills theory 
with $24$ matter fields in the adjoint representation of the $U(N)$ 
gauge group.

Then, by the conventional quantum field theory, the
$g_s^n$-order scattering amplitude $A$ can be extracted from the S-matrix
element described as a correlation function of $n$ interaction vertices $V(z_i)$ with 
the subsequent integration over the insertion points $z_i$:  
$$ 
\l f|S|i \r\sim \int
\prod_{i}d^2z_i \l f|V(z_1)\ldots V(z_n)|i \r.
$$

The paper is organized as follows. In the second section we remind the
description of the Hilbert space of the orbifold model. In the third section
the vertex operators that create the states of the Hilbert space are 
introduced and their conformal dimensions are calculated. We pay
special attention to the fact that in nonabelian orbifold models
there is no decomposition of vertex operators into the tensor
product of bosonic and fermionic (holomorphic and antiholomorphic)
twist fields. We also recall the construction of the DVV interaction 
vertex. In the fourth section we describe the S-matrix element
corresponding to the scattering of four gravitons and reduce
the problem of it's calculation to the one of computing 
special correlation functions in the orbifold CFT. In Section 5
we compute the bosonic correlation functions up to 
normalization constants by using the stress-energy tensor method. 
To compute the fermionic contribution, 
in the next section we describe a bosonization procedure for fermions 
of the orbifold model in the $SU(4)\times U(1)$ formalism.
In Section 7 we find normalization constants for the 
correlation functions. Finally, in Section 8 we combine the results 
obtained in the previous sections and derive the well-known
four graviton scattering amplitude that appears to be automatically
Lorentz-invariant. In Conclusion we discuss unsolved problems.

\section{$S^N\R^{8}$ supersymmetric orbifold sigma model}
\setcounter{equation}{0}
The target space of the supersymmetric orbifold sigma model
is the symmetric product space $S^N\R^{8}=(\R^8)^N/S_N$,
where $S_N$ is the permutation group of $N$ objects.
The model on a cylinder with coordinates $(\sigma,\tau)$ is described
by the following action
\be S=\frac{1}{2\pi}\int d\tau d\sigma
(\partial_{\tau}X^i_I\partial_{\tau}X^i_I
-\partial_{\sigma}X^i_I\partial_{\sigma}X^i_I +
\frac{i}{2}\theta^{a}_I(\partial_{\tau}+\partial_{\sigma})\theta^{a}_I+
\frac{i}{2}\theta^{\a}_I(\partial_{\tau}-\partial_{\sigma})\theta^{\a}_I),
\label{act}
\ee
Here $0\leq \sigma < 2\pi$, $I=1,2,\dots ,N$.
The real bosonic fields $X^i$, $i=1,2,\dots ,8$ transform in the ${\bf 8_v}$
representation of the $SO(8)$ group, while the components
$\theta^{a},\theta^{\a}$, $a,\a=1,\dots ,8$
of the 16-component Majorana-Weyl spinor $\theta^{\alpha}$
transform in the ${\bf 8_s}$ and ${\bf 8_c}$ representations respectively.
One has also to identify
all configurations $(X,\theta)$ related by arbitrary $S_N$ transformations:
\be
\la{iden}
X\sim hX,~~~\theta\sim h\theta, \qquad h\in S_N.
\ee

As usual in orbifold models \cite{DHVW1,DHVW2}, the fields $X^i,\theta^{\alpha}$
can have twisted boundary conditions
\be
X^i(\sigma +2\pi )=gX^i(\sigma ),\qquad
\theta^{\alpha}(\sigma +2\pi )=g\theta^{\alpha}(\sigma ), \qquad g\in S_N.
\label{bc}
\ee
Note that the untwisted sector corresponds to the Ramond boundary condition.

Multiplying (\ref{bc}) by some element $h\in S_N$ and taking into
account the identification (\ref{iden}), one gets that all
possible boundary conditions are in one-to-one correspondence with the
conjugacy classes of the symmetric group. Therefore, the Hilbert space of the
orbifold model is decomposed into the direct sum of Hilbert spaces of the
twisted sectors corresponding to the conjugacy classes $[g]$ of $S_N$
\cite{DMVV}
\bea
\cH(S^N\R^{D}) = \bigoplus_{[g]} \cH_{[g]}.
\nonumber
\eea
It is well-known that the conjugacy classes of $S_N$ are described by
partitions $\{ N_n\}$ of $N$
\bea
N=\sum_{n=1}^s\, nN_n
\nonumber
\eea
and can be represented as
\be
[g] = (1)^{N_1}(2)^{N_2} \cdots (s)^{N_s}.
\label{facg}
\ee
Here $N_n$ is the multiplicity of the cyclic permutation $(n)$
of $n$ elements.

In any conjugacy class $[g]$ there is the only element $g_c$ that has  the
canonical block-diagonal form
\bea
\nonumber
g_c=diag(\underbrace{\omega _1,...,\omega _1}_{N_1\ times},
\underbrace{\omega _2,...,\omega _2}_{N_2\ times},...,
\underbrace{\omega _s,...,\omega _s}_{N_s\ times}),
\eea
where $\omega _n$ is an $n\times n$ matrix that generates the cyclic
permutation $(n)$ of $n$ elements
\bea
\omega _n=\sum _{i=1}^{n-1}E_{i,i+1} +E_{n1}
\nonumber
\eea
and $E_{ij}$ are matrix unities.

\noindent It is not difficult to show that $\omega _n$ generates the $\Z_n$
group, since $\omega _n^n=1$, and that only the matrices $\omega _n^k$ from
$\Z_n$ commute with $\omega _n$.
Since the centralizer subgroup $C_g$ of any element $g\in [g]$ is isomorphic
to $C_{g_c}$ one concludes that
\bea
C_g =
\prod_{n=1}^s \, S_{N_n} \times \Z^{N_n}_n,
\nonumber
\eea
where the symmetric group $S_{N_n}$ permutes the $N_n$ cycles $(n)$. It is
obvious that the centralizer $C_g$ contains $\prod_{n=1}^s \, N_n!n^{N_n}$
elements.

Due to the factorization (\ref{facg}) of $[g]$, the Hilbert space
$\cH_{[g]}\equiv\cH_{\{ N_n\}}$ of each twisted sector can be decomposed into
the graded $N_n$-fold symmetric tensor products of the Hilbert spaces
$\cH_{(n)}$ which correspond to the cycles of length $n$
\bea
{\cH}_{\{N_n\}} = \bigotimes_{n=1}^s \, S^{N_n} \cH_{(n)}
=\bigotimes_{n=1}^s \,\left(\underbrace{\cH_{(n)} \otimes \cdots
\otimes \cH_{(n)}}_{N_n\ times}\right)^{S_{N_n}}.
\nonumber\eea
The space $\cH_{(n)}$ is $\Z_n$ invariant subspace of the Hilbert space of a
sigma model of $8n$ bosonic fields $X_I^i$ and $16n$ fermionic fields
$\theta^{\alpha}$ with the cyclic boundary condition
\be
X_I^i(\s +2\pi )=X_{I+1}^i(\s ), \qquad
\theta_I^{\alpha}(\s +2\pi )=\theta_{I+1}^{\alpha}(\s ),
\quad I=1,2,...,n.
\label{cyc}
\ee
The fields $X_I(\s )$ $(\theta_I(\s ))$ can be glued together into one field $X(\s )$
$(\theta(\s ))$ that is identified with a long string of the length $n$.
The states of the space $\cH_{(n)}$ are obtained by acting by the creation operators of the string on
eigenvectors of the momentum operator. These eigenvectors have the standard
normalization
\bea
\langle \q |\k\rangle =\d ^D(\q +\k )
\nonumber\eea
and can be regarded as states obtained by acting by the operator $\e ^{i\k x}$ on
the vacuum state \footnote{As is discussed below, the vacuum state carries
a representation of the Clifford algebra, since the long string is in
the Ramond sector.} (that is not normalizable):
$|\k\rangle =\e ^{i\k x}|0\rangle$, $\langle \q |=\langle 0 |\e ^{i\q x}$.

The $\Z_n$ invariant subspace  is
singled out by imposing the condition
\bea
(L_0-\bar {L}_0)|\Psi\rangle =nm|\Psi\rangle ,
\nonumber\eea
where $m$ is an integer and $L_0$ is the canonically normalized $L_0$ operator
of the single string.

The Fock space of the second-quantized IIA type string is
recovered in
the limit $N\to\infty$, $\frac {n_i}{N}\to p^+_i$ \cite{DMVV}, where the finite
ratio  $\frac {n_i}{N}$ is identified with the $p^+_i$ momentum of a long
string. The $\Z_n$ projection reduces in this limit to the usual level-matching
condition $L_0^{(i)}-\bar {L}_0^{(i)}=0$. The individual $p^-_i$ light-cone
momentum is defined by means of the standard mass-shell condition
$p^+_ip^-_i=L_0^{(i)}$.

\section{Vertex operators}
\setcounter{equation}{0}
Let us consider conformal field theory of free fields described by
the action (\ref{act}).
It is convinient to perform the Wick rotation
$\tau\to -i\tau$ and to map the cylinder onto the sphere:
$z=\e ^{\tau +i\s }$, $\bar {z}=\e ^{\tau -i\s }$.

The NS vacuum state $|0\rangle$ of the CFT is annihilated by the 
momentum operators and by annihilation operators, and has to be 
normalizable. To be able to identify this vacuum state with the 
vacuum state of the untwisted sector of the orbifold sigma model we 
choose the following normalization of $|0\rangle$ 
\bea \langle 0 
|0\rangle =R^{8N}.  
\nonumber\eea 
Here $R$ should be 
regarded as a regularization parameter of the sigma model.  We 
regularize the sigma model by compactifying the coordinates $x^i_I$ 
on circles of radius $R$. Then the norm of the eigenvectors of the 
momentum operators in the untwisted sector is given by 
\bea \langle 
\q |\k\rangle =(2\pi )^{-8N}\int_0^{2\pi R}d^{8N}x \e ^{i(\q +\k )x}= 
\prod_{I=1}^N \d ^8_R(\q_I +\k_I ),
\nonumber\eea
where $k^i_I=\frac {m^i_I}{R}$ and $q^i_I=\frac {n^i_I}{R}$ are momenta of the
states, $m^i_I$ and $n^i_I$ are integers since we compactified the coordinates,
and $\d ^8_R(\k )=R^8\prod_{i=1}^8\d _{m^i0}$ is the regularized $\d$-function.
In the limit $R\to\infty$ one recovers the usual normalization of the
eigenvectors.

The asymptotic states of the orbifold
CFT model should be created by some vertex operators applied to the
NS vacuum $|0\r$. Evidently, the vertex operators creating
the ground states of twisted sectors are in one-to-one correspondence
with the conjugacy classes of $S_N$. For nonabelian groups
a conjugacy class $[g]$ of an element $g$ contains many group
elements. It enforces us to define a vertex operator $V_{[g]}$
in two steps. First one introduces vertex operators $V_g$
corresponding to elements $g\in S_N$. The fundamental fields
obey the twisted boundary condition (\ref{bc}) around an insertion
point of an operator $V_g$. Under the group action, $V_g$ transforms
into $V_{h^{-1}gh}$ and, therefore, to define an invariant
operator $V_{[g]}$ one should sum up all vertex operators from
a given conjugacy class:
\bea
V_{[g]}(z,\bar z)=\frac {1}{N!}\sum_{h\in S_N} V_{h^{-1}gh}(z,\bar z).
\nonumber
\eea
Vertex operators creating excited states of the twisted sectors
can be defined in an analogous way.
The main requirement imposed on all invariant vertex operators is
that they should form a closed operator algebra.
Schematically the OPE of any noninvariant vertex
operators is of the form
\bea
\la{OPE}
V_{g_1}(z,\bar{z})V_{g_2}(0)=\frac{1}{z^{\Delta}\bar{z}^{\bar{\Delta}}}
\left(C_{g_1,g_2}^{g_1g_2}V_{g_{1}g_{2}}(0)+
C_{g_1,g_2}^{g_2g_1}V_{g_{2}g_{1}}(0)\right)+\cdots ,
\eea
where $\Delta,\bar{\Delta}$ are defined by the conformal
symmetry. Here the two leading terms appear because there are two
different ways to go around the points $z$ and $0$. It is not
difficult to see that $g_1g_2$ and $g_2g_1$ belong to the same
conjugacy class and, hence, $\D _{g_1g_2}=\D _{g_2g_1}$.
If one requires the operator algebra of invariant operators
to be closed, then one faces hard restrictions on the structure
constants in (\ref{OPE}). In particular, the structure constants
occuring in the OPE of operators $V_g$ creating
ground states should be invariant with respect to the global
action of $S_N$, e.g.:
$$
C_{h^{-1}g_1h,h^{-1}g_2h}^{h^{-1}g_1g_2h}=C_{g_1,g_2}^{g_1g_2}.
$$
Then OPE (\ref{OPE}) leads to the following OPE for invariant ground
state operators:
\bea
\nonumber
V_{[g_1]}(z,\bar{z})V_{[g_2]}(0)=
\frac{1}{N!}\sum_{h\in
S_N}\frac{1}{z^{\Delta_h}\bar{z}^{\bar{\Delta}_h}}
\left(C_{g_1,h^{-1}g_2h}^{g_1h^{-1}g_2h}+
C_{g_1,h^{-1}g_2h}^{h^{-1}g_2hg_1}\right)V_{[g_{1}h^{-1}g_{2}h]}(0)+\cdots ,
\eea
i.e., for leading terms the operator algebra is closed.

Naively, one can think that vertex operators $V_g$
receiving contributions from bosons and fermions can be
decomposed into tensor product of bosonic and
fermionic twist fields:
$V_g=\s_g \otimes\S_g$. Obviously, the OPE for the
fields $\s_g$ and $\S_g$ should be of the same type as for
$V_g$ (\ref{OPE}). Then one can easily see
that the tensor product structure of $V_g$
leads to the appearance of unwanted terms
in the OPE:
\bea
\nonumber
V_{g_1}V_{g_2}&=&(\s_{g_1} \otimes\S_{g_1})(\s_{g_2} \otimes\S_{g_2})=
\s_{g_1}\s_{g_2}\otimes \S_{g_1}\S_{g_2}\\
\nonumber
&\sim &
\left(B_{g_1,g_2}^{g_1g_2}\s_{g_{1}g_{2}}+
B_{g_1,g_2}^{g_2g_1}\s_{g_{2}g_{1}}\right)\otimes
\left(F_{g_1,g_2}^{g_1g_2}\S_{g_{1}g_{2}}+
F_{g_1,g_2}^{g_2g_1}\S_{g_{2}g_{1}}\right)+\cdots \\
\nonumber
&=&
B_{g_1,g_2}^{g_1g_2}F_{g_1,g_2}^{g_1g_2}V_{g_{1}g_{2}}
+B_{g_1,g_2}^{g_2g_1}F_{g_1,g_2}^{g_2g_1}V_{g_{2}g_{1}} +\\
\nonumber
&&
B_{g_1,g_2}^{g_1g_2}F_{g_1,g_2}^{g_2g_1}\s_{g_{1}g_{2}}\otimes
\S_{g_{2}g_{1}}+
B_{g_1,g_2}^{g_2g_1}F_{g_1,g_2}^{g_1g_2}\s_{g_{2}g_{1}}\otimes
\S_{g_{1}g_{2}} + \cdots .
\eea
The same arguments also reveal the absence of decomposition of $V_g$ into
the tensor product of holomorphic and antiholomorphic parts.  A
reason for the absence of tensor product structure lies, of course, in
the nonabelian nature of the $S_N$ orbifold CFT. However,
in what follows to simplify the notation we represent $V_g$
as a product of bosonic and fermionic (holomorphic and antiholomorphic)
twist fields:
$V_g(z,\bar{z})=\s_{g}(z)\S_{g}(z)\bar{\s}_{g}(\bar{z})\bar{\S}_{g}(\bar{z})$.

It is known that any $g\in S_N$ has the decomposition
\be
(n_1)(n_2)\cdots (n_{N_{str}}),
\la{facg1}
\ee
where each cycle of length $n$ has
a definite set of indices ordered up to a cyclic permutation and
generates the action of the subgroup ${\bf Z}_n$.
Due to this decomposition the vertex operator $V_g$ can be represented
as the following product
\bea
V_g=\prod_{\al =1}^{N_{str}} \, V_{(n_\al )},
\nonumber
\eea
where $V_{(n)}$ is a vertex operator
that creates the vacuum state of the space $\cH_{(n)}$ of the sigma model
of fundamental fields with cyclic boundary condition (\ref{cyc}).

It is obvious that the conformal dimensions of the vertex operators
corresponding to cycles of the same length coincide\footnote{It explains
why we do not specify a set of indices occured in a given cycle.}
and therefore $\D _g$ depends only on $[g]$ and is given by the
equation
\be
\D _g=\sum_{\alpha =1}^{N_{str}}\Delta_{n_{\al }}=\sum_{n=1}^s\, N_n\D _{n},
\label{confdim}
\ee
where $\D _{n}$ denotes the
conformal dimension of the vertex operator $V_{(n)}$.
Thus, it is enough to consider the operator $V_{(n)}$, which we again
represent as a product of bosonic and fermionic
twist fields.

We begin with describing the bosonic twist operator.
As usual, the field $X(z,\bar {z})$ can be decomposed into the left- and
right-moving components
\be
2X(z,\bar {z})=X(z)+\bar {X}(\bar {z}).
\label{xdec}
\ee
In what follows we shall mainly concentrate our attention on the left-moving
sector.

Let $\s _{(n)}(z,\bar{z})$ be a primary field \cite{BPZ} that creates
a bosonic vacuum of the twisted sector at the point $z$, i.e.
the fields $X^i(z)$ satisfy the following monodromy
conditions
\bea
X^i(z\e^{2\pi i}, \bar{z}\e^{-2\pi i})\s _{(n)}(0) =\omega_n
X^i(z,\bar{z})\s _{(n)}(0),
\la{mon}
\eea
where $\omega_n$ generates the cyclic permutation of $n$ elements.

It is also convinient to regard
the twist field $\s _{(n)}(z,\bar{z})$ as a
product $\s _{(n)}(z,\bar{z})=\s _{(n)}(z)\bar{\s}_{(n)}(\bar{z})$.
To simplify calculations, we
require that under the world-sheet parity transformation
$z\to \bar{z}$, and $X(z)\to \bar{X}(\bar{z})$ the field
$\s _{(n)}(z)$ transforms into $\bar{\s}_{(-n)}(\bar{z})$, where $(-n)$ denotes
the cycle with the reversed orientation corresponding to the element
$\omega_n^{-1}$.

Note that (\ref{mon}) does not completely specify the field $\s_g$ but it
contains enough information to derive it's conformal dimension.
For later use we consider the general case of $Dn$ bosonic
fields.

Let the twist field $\s _{(n)}$ be located at $z=0$ and
let us denote the vacuum
state \footnote{This vacuum state is a primary state of the CFT.} as
$|(n)\rangle =\s _{(n)}(0,0)|0\rangle$.  Since the twist field $\s
_{(n)}$ creates one long string we normalize the vacuum state
$|(n)\rangle$ as \be \langle (n) |(n)\rangle =R^{D}.
\label{normvac1}
\ee

The fields $X(z)$ have the following
decomposition in the vicinity of $z=0$
\be
\partial X_I^i(z)=-\frac i n \sum_m \al _m^i\e^{-\frac {2\pi i}{n}Im}
z^{-\frac {m}{n}-1},
\label{moddec}
\ee
where $\al _m^i$ ($m\neq 0$) are the usual creation and annihilation operators
with the commutation relations
\be
[\al _m^i, \al _n^j]=m\d ^{ij}\d _{m+n,0},
\label{comrel}
\ee
and $\al _0^i$ is proportional to the momentum operator
\footnote{$\al _0^i=\frac{1}{2}p^i$ in string units $\alpha'=\frac{1}{2}$.}.

\noindent The vacuum state $|(n)\rangle $ is annihilated by the operators
$\al _m^i$ for $m\geq 0$.

Since $\s _{(n)}$ is a primary field, the conformal dimension $\D^b_{n}$ can be
found from the equation
\bea
\langle (n)|T(z)|(n)\rangle =\frac {\D^b_{n}}{z^2}\langle (n)|(n)\rangle ,
\nonumber
\eea
where $T(z)$ is the stress-energy tensor.

\noindent By using eqs. (\ref{moddec}) and (\ref{comrel}), one calculates the
correlation function
\bea
\langle (n)|\partial X_I^i(z)\partial X_I^j(w)|(n)\rangle
=-\d ^{ij}\frac {(zw)^{\frac 1 n -1}}{n^2(z^{\frac 1 n} -w^{\frac 1 n})^2}
\langle (n)|(n)\rangle .
\nonumber
\eea
Taking into account that the stress-energy tensor is defined as
\bea
T(z)= -\frac 12 \lim _{w\to z}\sum_{i=1}^{D}\sum_{I=1}^n\left(
\partial X_I^i(z)\partial X_I^i(w) +\frac {1}{(z-w)^2}\right) ,
\nonumber
\eea
one gets
\be
\D^b_{n}=\frac {D}{24} (n-\frac {1} {n} ).
\label{confdim1}
\ee

The excited states of this sigma model are obtained by acting on $|(n)\rangle $
by some vertex operators. In particular the state corresponding to a scalar
particle with momentum $\k$ is given by
\be
\s _{(n)}[\k](0,0)|0\rangle =:\e ^{ik^i_IX_I^i(0,0)}:|(n)\rangle ,
\label{tachst}
\ee
where the summation over $i$ and $I$ is assumed, $k^i_I=\frac {m^i_I}{R}$ is a momentum carried
by the field $X_I^i(z,\bar z)$ and $k^i=\sum_{I=1}^n k^i_I$ is a total momentum
of the long string.

\noindent By using the definition of the vacuum state $|(n)\rangle$, one can
rewrite eq.(\ref{tachst}) in the form
\be
\s _{(n)}[\k ](0,0)|0\rangle =:\e ^{i\frac {k^i}{\sqrt {n}}Y^i(0,0)}:
|(n)\rangle ,
\label{tachst1}
\ee
where
\be
Y^i(z,\bar z)=\frac {1}{\sqrt {n}}\sum_{I=1}^n X_I^i(z,\bar z).
\label{defY}
\ee
The field $Y(z)$ is canonically normalized, i.e. the part of the stress-energy
tensor depending on $Y$ is $-\frac 12 :\partial Y(z)\partial Y(z):$, and has
the trivial monodromy around $z=0$.

It is obvious from eq.(\ref{tachst1}) that the conformal dimension of the
primary field
$$
\s _{(n)}[\k ](z,\bar z) =
:\e ^{i\frac {k^i}{\sqrt {n}}Y^i(z,\bar z)}:\s _{(n)}(z,\bar z)
$$
is equal to
\bea
\D^b_{n}[\k ]=\D^b_{n}+\frac {\k ^2}{8n}=
\frac {D}{24} (n-\frac 1n)+\frac {\k ^2}{8n},
\nonumber
\eea
where the decomposition (\ref{xdec}) was taken into account.

To simplify calculations it will be convinient to treat $\s _{(n)}[\k
](z,\bar z)$ as a product of holomorphic and antiholomorphic parts
$\s _{(n)}[\k /2 ](z)\s _{(n)}[\k /2](\bar z)$.

Other excited states of the model can be produced by considering
the OPE of the fields $\partial X$ with the twist fields.
By using (\ref{moddec}) and the definition of the vacuum state
$|(n)\r$ one can see that the most singular term of the OPE looks as
\be
\la{tau}
\partial X_I^i(z) \s_{(n)}
(w,\bar{w})=(z-w)^{-\left(1-\frac{1}{n}\right)}e^{\frac{2\pi i}{n}I}
\tau_{(n)}^i(w,\bar{w})+\ldots,
\ee
where $\tau_{(n)}^i(0,0)=-\frac{i}{n}\al^i_{-1}|(n)\r$ is
the first excited state in the twisted sector.
According to our conventions, the field
$\tau_{(n)}^i(z,\bar{z})$ can be represented
as a product:
$\tau_{(n)}^i(z,\bar{z})=\tau_{(n)}^i(z)\bar{\s}_{(n)}(\bar{z})$.
In particular, since the element
$g_{IJ} =1-E_{II}-E_{JJ}+ E_{IJ}+E_{JI}$ transposing the fields $X_I$
and $X_J$ has just one cycle of length $2$, one can define the field
$\t_{IJ}\equiv \t_{(2)}$. This twist field will be used to define
the DVV interaction vertex.

Similarly to the vertex operator $V_g$, the twist field $\s_g$ can be
represented as
\bea
\s _g=\prod_{\al =1}^{N_{str}} \, \s _{(n_\al )}.
\nonumber
\eea

Due to eqs. (\ref{confdim}) and (\ref{confdim1}), the
conformal dimension of $\s _g$ is given by
\bea
\D _g=\sum_{n=1}^s\,
N_n\frac {D}{24} (n-\frac 1n)= \frac {D}{24}(N-\sum_{n=1}^s\, \frac
{N_n}{n}).
\label{confdim2}
\nonumber
\eea
One can also introduce a
primary field that creates scalar particles with momenta $k^i_\al$,
$\al =1,2,...,N_1+N_2+\cdots +N_s\equiv N_{str}$
\bea
\s _{g}[\{\k _\al\}](z,\bar z) = :\e ^{i\frac {k^i_\al }{\sqrt {n_\al}}Y^i_\al
(z,\bar z)}:\s _{g}(z,\bar z)=\prod_{\al =1}^{N_{str}} \,
\s_{(n_\al )}[\k_\al ], 
\nonumber
\eea
where $n_1=n_2=\cdots =n_{N_1}=1$, $n_{N_1+1}=n_{N_1+2}=\cdots
=n_{N_1+N_2}=2$ and so on, $Y^i_\al $ corresponds to the cycle
$(n_\al )$ and is defined by eq.(\ref{defY}), and the summation over
$i$ and $\al$ is assumed.

\noindent The conformal dimension of the field $\s _{g}[\{\k _\al\}]$ is equal
to
\bea
\D _g[\{\k _\al\}]=
\frac {D}{24}(N-\sum_{n=1}^s\, \frac {N_n}{n})+\sum_\al
\frac {\k^2_{\al }}{8n_{\al} }.
\label{confdim3}
\eea
It is obvious that the two-point correlation function of the twist fields
$\s_{g_1}$ and $\s_{g_2}$ is not equal to zero if and only if $g_1g_2=1$.
Taking into account the
normalization (\ref{normvac1}), we find
\footnote{it is clear that $[g^{-1}]=[g]$ and therefore $\D _{g^{-1}}=\D _{g}$}
\bea
\langle \s_{g^{-1}}(\infty)\s_g(0)\rangle =R^{DN_{str}}.
\nonumber
\eea
It means that the fields $\s_{g^{-1}}$ and $\s_g$ have the following OPE
\bea
\s_{g^{-1}}(z,\bar z)\s_g(0,0)=
\frac {R^{D(N_{str}-N)}}{|z|^{4\D _{g}}}+\cdots .
\nonumber
\eea
Here we assume that $\l 0|0\r=R^{DN}$.

The two-point correlation function of $\s _{g^{-1}}[\{\q _\al\}]$ and
$\s _{g}[\{\k _\al\}]$ is respectively equal to
\be
\langle \s_{g^{-1}}[\{\q _\al\}](\infty)\s_g[\{\k _\al\}](0)\rangle =
\prod_\al  \d ^D_R(\q_\al +\k_\al ).
\label{norm}
\ee

Now we proceed with describing the fermion twist fields
that create the vacuum states corresponding to long strings.
We begin with the case of one long string of length $n$.

Equations of motion
corresponding to the cyclic boundary condition (\ref{cyc}) imply that
$\theta^{a}$ and $\theta^{\a}$ are the following holomorphic and
antiholomorphic functions on the $z$-plain:
\bea
\la{fer}
\theta^{a}_I(z)=\frac{1}{\sqrt{n}}\sum_m \theta_m^a e^{-\frac{2\pi i}{n}Im}z^{-\frac{m}{n}-\frac{1}{2}},\\
\nonumber
\theta^{\a}_I(\bar{z})=\frac{1}{\sqrt{n}}\sum_m \theta_m^{\a} e^{\frac{2\pi i}{n}Im}\bar{z}^{-\frac{m}{n}-\frac{1}{2}},
\eea
where we have taken into account that under conformal mappings fermion
fields have the scaling dimension $1/2$.
Thus, on the $z$-plain the fermions satisfy $\theta^a_I(ze^{2\pi i})=-\theta^a_{I+1}(z)$
and analogously for $\theta^{\a}_I$. In what follows
we mainly concentrate on the left-moving sector.

\noindent In eq.(\ref{fer}) the creation and annihilation operators
satisfy the standard commutation relations:
\be
\la{ca}
\{\theta^a_m,\theta^b_n\}=\d^{ab}\d_{m+n},
\ee
i.e. zero modes $\theta^a_0$ form the Clifford algebra. Therefore,
the vacuum state anihilated by $\theta^a_m$ for
$m > 0$ carry an irreducible representation of the Clifford algebra. By the
triality the representation space can be choosen as the direct sum
${\bf 8_v}+{\bf 8_c}$. It means that the vacuum state is a 16-component vector
with components $|i\rangle$ and $|\a\rangle$ normalized in the standard
fashion $\l i|j\r=\d^{ij}$, $\l\a|\b\r=\d^{\a\b}$ and
transforming under the action
of $\theta^{a}_0$ as follows
\bea
\la{tha}
\theta^{a}_0|i\r=\frac{1}{\sqrt{2}}\g^i_{a\a}|\dot{a}\r,\qquad
\theta^{a}_0|\dot{a}\r=\frac{1}{\sqrt{2}}\g^i_{a\a}|i\r.
\eea
In the CFT the vacuum states $|i\rangle$ and $|\a\rangle$ are created by
the primary  (spin) fields $\Sigma^i_{(n)}$ and $\Sigma^{\a}_{(n)}$. Their
conformal dimension can be found similarly to the bosonic case.
Denoting by $\Sigma_{(n)}^{\dot{\mu}}$ one of the
fields $\Sigma^i_{(n)}$, $\Sigma^{\a}_{(n)}$ and using eq.(\ref{ca})
we obtain
\bea
\nonumber
\l\Sigma_{(n)}^{\dot{\mu}}|\theta^{a}_I(z)\partial\theta^{b}_I(w)|\Sigma_{(n)}^{\dot{\mu}}\r =
-\frac{1}{2}\frac{\l\Sigma_{(n)}^{\dot{\mu}}|\theta^a_0\theta^b_0|\Sigma_{(n)}^{\dot{\mu}}\r}{nz^{1/2}w^{3/2}}
+\frac{\d^{ab}}{nz^{1/2}}\left(
\frac{\left(\frac{1}{n}-\frac{1}{2}\right)w^{1/n-3/2}}{z^{1/n}-w^{1/n}}+
\frac{\frac{1}{n}w^{2/n-3/2}}{(z^{1/n}-w^{1/n})^2}
\right).
\eea
Taking into account the definition of the stress-energy tensor $T^F$
for fermion fields
$$
T^F(z)=-\frac{1}{2}\lim_{w\to z}\sum_{a=1}^8\sum_{I=1}^n
\left(\theta^{a}_I(z)\partial\theta^{b}_I(w)-\frac{1}{(z-w)^2}
\right),
$$
one finds
\be
\la{fd}
\Delta^f_n=\frac{n}{6}+\frac{1}{3n}.
\ee
The transformation properties (\ref{tha}) are encoded in the following OPE:
\be
\la{fOPE}
\theta^{a}_I(z)\Sigma^i_{(n)}(0)=\frac{1}{\sqrt{n}z^{1/2}}\frac{\g^i_{a\a}}{\sqrt{2}}\Sigma^{\a}_{(n)}(0)
+\ldots,
\qquad
\theta^{a}_I(z)\Sigma^{\a}_{(n)}(0)=\frac{1}{\sqrt{n}z^{1/2}}\frac{\g^i_{a\a}}{\sqrt{2}}\Sigma^i_{(n)}(0)+\ldots
\ee
Note that the r.h.s. of this OPE contains other (less) singular terms
that correspond to the excited states of the twisted sector.

The twist fields $\bar{\S}^{\mu}_{(n)}(\bar{z})$ for the right-moving
sector are introduced in the same manner, they realize the
representation space ${\bf 8_v}+{\bf 8_s}$, and have the same
conformal dimensions.
The fields $\theta^{\a}(\bar{z})$ have the following OPE with the
twist fields
\be
\la{fOPEa}
\theta^{\a}_I(\bar{z})\bar{\Sigma}^i_{(n)}(0)
=-\frac{1}{\sqrt{n}\bar{z}^{1/2}}
\frac{\g^i_{a\a}}{\sqrt{2}}\bar{\Sigma}^{a}_{(n)}(0)
+\ldots,
\qquad
\theta^{\a}_I(\bar{z})\bar{\Sigma}^{a}_{(n)}(0)
=-\frac{1}{\sqrt{n}\bar{z}^{1/2}}
\frac{\g^i_{a\a}}{\sqrt{2}}\bar{\Sigma}^i_{(n)}(0)+\ldots
\ee
Comparing eqs.(\ref{fOPE}) and (\ref{fOPEa}) one can see that under
the world-sheet parity transformation $z\to \bar{z}$ and the space
reflection $X^3\to -X^3$ the fermions and twist fields transform
as follows:
\bea
\nonumber
&&\theta^a(z)\leftrightarrow \theta^{\a}(\bar{z}); \qquad
\Sigma^{\a}_{(n)}(z)\leftrightarrow\bar{\Sigma}^{a}_{(-n)}(\bar{z});\\
\label{trans}
&&\Sigma^i_{(n)}(z)\leftrightarrow\bar{\Sigma}^i_{(-n)}(\bar{z}), \qquad
i\neq 3; \qquad  \Sigma^3_{(n)}(z)\leftrightarrow
-\bar{\Sigma}^3_{(-n)}(\bar{z}).
\eea
The third direction is singled out since in our conventions
$\g^3=1$ (see Appendix A).

At last combining the fermionic vacuum states of the holomorphic and the antiholomorphic
sectors with the vacuum state of the bosonic sector we obtain 256 states that
describes the spectrum of the IIA supergravity.
The IIA supergravity states with the momentum $\k$
are
\bea
\nonumber
|V(\k,\dot{\mu},\nu)\r=|\k\r \otimes |\dot{\mu}\r\otimes |\nu\r,
\eea 
where $\dot{\mu}=(i,\a)$ and $\mu=(i,a)$.  Clearly, these
states can be generated from the NS vacuum $|0\r $ by the following
vertex operators:
\bea 
\nonumber
V_{(n)}[\k,\dot{\mu},\nu](z,\bar{z})=
\s_{(n)}[\k](z,\bar{z}) \Sigma^{\dot{\mu}}_{(n)}(z)
\bar{\Sigma}^{\nu}_{(n)}(\bar{z}).
\eea
The conformal dimension of the vertex operator is equal to
\be
\la{cdv}
\Delta_n=\frac{n}{2}+\frac{\k^2}{8n}.
\ee
In particular, a graviton with a momentum $\k$ and a polarization
$\z$ is created by
\bea 
\nonumber
V_{(n)}[\k,\z](z,\bar{z})=\z_{ij}
\s_{(n)}[\k](z,\bar{z}) \Sigma^{i}_{(n)}(z)
\bar{\Sigma}^{j}_{(n)}(\bar{z}),
\eea
where $\z_{ij}$ is a symmetric tensor.

It is worth noting that due to eq.(\ref{trans}) under
the world-sheet parity transformation $z\to \bar{z}$ and the space
reflection $X^3\to -X^3$ the graviton
vertex operator $V_{(n)}[\k,\z]$ transforms into
$V_{(-n)}[\tilde{\k},\tilde{\z}]$, where $\tilde{k}^i,\tilde{\z}_{ij}$
are the space reflected momenta and polarizations respectively
$(\tilde{k}^3=-k^3)$.

Due to the factorization (\ref{facg1}) the vertex operator
corresponding  to any element $g\in S_N$ has the following
decomposition into the tensor
product of $V_{(n)}[\k,\dot{\mu},\nu]$:
\bea
\nonumber
V_{g}[\{\k _\al,\dot{\mu}_{\al},\nu_{\al}\}]
=\prod_{\al =1}^{N_{str}} \,
V_{(n_\al )}[\k _\al,\dot{\mu}_{\al},\nu_{\al}].
\eea
According to (\ref{cdv}) the conformal dimension  of $V_{g}$
is given by
\be
\la{cdver}
\Delta_g=\frac{N}{2}+\sum_{\al =1}^{N_{str}}\frac{\k_{\al}^2}{8n}.
\ee
Thus, we see that the operators creating ground states ($\k_{\al }=0$)
have the same conformal dimension that does not depend on a
particular group element.

As was discussed before an invariant vertex operator
is defined by summing up all the twist fields from
one conjugacy class:
\bea
V_{[g]}[\{\k _\al,\dot{\mu}_{\al},\nu_{\al}\}]
=\frac {1}{N!}\sum_{h\in S_N}\prod_{\al =1}^{N_{str}} \,
V_{h^{-1}(n_\al )h}[\k _\al,\dot{\mu}_{\al},\nu_{\al}].
\label{invver}
\eea
One can easily check that the vertex operators are invariant
with respect to the simultaneous permutation of $\k _\al $,
$\dot{\mu}_{\al}$ and $\nu_{\al}$ which correspond to cycles
$(n_\al )$ of the same length.

By using this definition, one can easily calculate the two-point correlation
function
\bea
\l
V_{[g]}[\{\k _\al,\dot{\mu}_{\al},\rho_{\al}\}](\infty)
V_{[g]}[\{\q _\al,\dot{\nu}_{\al},\epsilon_{\al}\}](0)\r =
\frac {1}{N!}
\prod_{n=1}^sN_n!n^{N_n}
\prod_{\al}  \d ^8_R(\q_\al +\k_\al )
\d^{\dot{\mu}_{\al}\dot{\nu}_{\al}} \d^{\rho_{\al}\epsilon_{\al}},
\nonumber\eea
where $\prod_{n=1}^sN_n!n^{N_n}$ is the number of elements of the centralizer
subgroup $C_g$.

Thus, we have introduced the vertex operators that create asymptotic
states corresponding to the massless particles of the type IIA string.

To describe the interaction vertex proposed by DVV \cite{DVV}
we need another kind of spin twist fields. Note that on the $z$-plane
there are twist fields around which fermions obey the following
boundary condition:
$$
\theta(e^{2\pi i}z)=g\theta(z).
$$
Consider the group element
$g_{IJ}$ transposing the fields $\theta_I$
and $\theta_J$. Since the combination $\theta_I-\theta_J$ satisfies
the Ramond boundary condition the corresponding spin field carries a
representation of the Clifford algebra.
The DVV interaction vertex is defined with the help of
the twist field $\S^i _{IJ}$ transforming as a vector of $SO(8)$.
In fact, $\S^i _{IJ}$ is a well-known spin field of the $\R^8/\Z_2$
supersymmetric orbifold sigma model.

To write down the DVV interaction vertex it is useful to come back to the
Minkowskian space-time. Then the interaction is described by the
translationally-invariant vertex
\bea
V_{int}=\frac{\lambda N}{2\pi } \sum_{I<J}\int d\tau d\s
\left(\t^i(\s _+)\S^i(\s _+)\bar{\t}^j(\s _-)\bar{\S}^j(\s _-)\right)_{IJ},
\nonumber
\eea
where $\lambda$ is a coupling constant proportional to the string coupling,
and $\s _\pm$ are light-cone coordinates: $\s _\pm =\tau\pm\s$.

The twist field $V_{IJ}(\s _+,\s _-)=\left(\t^i(\s _+)\S^i(\s _+)
\bar{\t}^j(\s _-)\bar{\S}^j(\s _-)\right)_{IJ}$
is a weight $(\frac {3}{2} ,\frac {3}{2} )$
conformal field and the coupling constant
$\lambda$ has dimension $-1$. As was shown in \cite{DVV} the
interaction vertex is space-time supersymmetric, $SO(8)$ invariant
and describes an elementary string interaction.
Another important property of the interaction vertex is the invariance
with respect to the world-sheet parity transformation $\s\to -\s$ and
an odd number of space reflections.

Performing again the Wick rotation and the conformal map onto the sphere,
one gets the following expression for $V_{int}$
\bea
V_{int}=-\frac{\lambda N}{2\pi } \sum_{I<J}\int d^2z|z| V_{IJ}(z,\bar z),
\nonumber\eea
where the minus sign appears because $V_{IJ}$ has conformal dimension
$(\frac {3}{2} ,\frac {3}{2} )$.

Thus, the action of the interacting $S^N\R^{8}$ sypersymmetric
orbifold sigma model is given by the sum
\bea S_{int}=S_0+V_{int}
\nonumber
\eea
In the next section we calculate the S-matrix element corresponding to
the scattering of two gravitons and show that the scattering
amplitude coincides with the type IIA string scattering amplitude.

\section{S-matrix element}
\setcounter{equation}{0}
The S-matrix element at the second order in the coupling constant $\lambda$ is
given by the standard formula of quantum field theory
\be
\langle f|S|i\rangle = -\frac {1}{2}\left(\frac{\lambda N}{2\pi }\right)^2
\langle f|\int d^2z_1d^2z_2|z_1||z_2|T\left(
\cL_{int}(z_1,\bar z_1)
\cL_{int}(z_2,\bar z_2)\right) |i\rangle ,
\label{matel}
\ee
where the symbol $T$ means the time-ordering: $|z_1|>|z_2|$, and
\bea
\cL_{int}(z,\bar z)=\sum_{I<J} V_{IJ}(z,\bar z).
\nonumber\eea
The initial state $|i\rangle$ describes two gravitons with momenta $\k_1$ and
$\k_2$, and polarizations $\z_1$ and $\z_2$,
and is created by the vertex operator $V_{[g_0]}[\k_1,\z_1;\k_2,\z_2]$:
$$
V_{[g_0]}[\k_1,\z_1;\k_2,\z_2](z,\bar{z})=
\frac{1}{N!}\sum_{h\in S_N}V_{h^{-1}(n_0)h}[\k_1,\z_1](z,\bar{z})
V_{h^{-1}(N-n_0)h}[\k_2,\z_2](z,\bar{z}).
$$
Namely,
\bea |i\rangle
=C_0V_{[g_0]}[\k_1,\z_1;\k_2,\z_2](0,0)|0\rangle .
\nonumber\eea
Here the element $g_0$ is taken in the canonical block-diagonal form
$$g_0=(n_0)(N-n_0),$$ where $n_0<N-n_0$.

The final state $\langle f|$ describes two gravitons with momenta $\k_3$ and
$\k_4$, and polarizations $\z_3$ and $\z_4$,
and is given by the formula (see \cite{BPZ})
\bea
\langle f|=C_\infty \lim_{z_\infty\to\infty}|z_\infty |^{4\D_\infty}
\langle 0|V_{[g_\infty]}[\k_3,\z_3;\k_4,\z_4](z_\infty ,\bar z_\infty ).
\nonumber\eea
The element $g_\infty$ has the canonical decomposition
$$g_\infty =(n_\infty )(N-n_\infty ),\quad n_\infty <N-n_\infty .$$
The constants $C_0$ and $C_\infty$ are chosen to be equal to
$$
C_0=\sqrt {\frac {N!}{n_0(N-n_0)}},\quad
C_\infty =\sqrt {\frac {N!}{n_\infty (N-n_\infty )}}$$
that guarantees the standard normalization of the initial and final states.

After the conformal transformation $z\to \frac {z}{z_1}$ eq.(\ref{matel})
acquires the form
\bea
\langle f|S|i\rangle &=&-\frac {1}{2}\left(\frac{\lambda N}{2\pi }\right)^2
\int d^2z_1d^2z_2|z_1||z_2||z_1|^{2\D_\infty -2\D_0 -6}\nonumber\\
&\times &\langle f|T\left( \cL_{int}(1,1)
\cL_{int}(\frac {z_2}{z_1},\frac {\bar z_2}{\bar z_1})\right) |i\rangle ,
\nonumber
\eea
where, according to (\ref{cdver}),
the conformal dimensions $\D_0$ and $\D_\infty$ of the
vertex operators
$V_{[g_0]}[\k_1,\z_1;\k_2,\z_2]$ and
$V_{[g_\infty]}[\k_3,\z_3;\k_4,\z_4]$ are given by
\bea
\D_0 &=&\frac
{N}{2}+ \frac{\k_1^2}{8n_0}+\frac{\k_2^2}{8(N-n_0)},\nonumber\\
\D_\infty &=&\frac {N}{2}+
\frac {\k_3^2}{8n_\infty}+ \frac {\k_4^2}{8(N-n_\infty )}.
\la{confdim4}
\eea
Let us introduce the light-cone momenta of the gravitons
\cite{DVV} taking into account the mass-shell condition for the
graviton states
\bea
k_1^+&=&\frac {n_0}{N},\quad
k_1^-k_1^+-\k_1^2\equiv -k_1^2 =0,\nonumber\\ k_2^+&=&\frac
{N-n_0}{N},\quad k_2^-k_2^+-\k_2^2\equiv -k_2^2  =0,\nonumber\\
k_3^+&=&-\frac {n_\infty}{N},\quad k_3^-k_3^+-\k_3^2\equiv -k_3^2
=0,\nonumber\\ k_4^+&=&-\frac {N-n_\infty}{N},\quad
k_4^-k_4^+-\k_4^2\equiv -k_4^2  =0.
\nonumber\eea
By using the light-cone momenta and the mass-shell condition, one can rewrite
(\ref{confdim4}) in the form
\bea
\D_0 &=&\frac{N}{2}+\frac {k_1^-+k_2^-}{8N},\nonumber\\
\D_\infty &=&\frac{N}{2}-\frac {k_3^-+k_4^-}{8N}.
\nonumber\eea
Performing the change of variables $\frac {z_2}{z_1}=u$, one obtains
\bea
\langle f|S|i\rangle &=& -\frac {1}{2}\left(\frac{\lambda N}{2\pi }\right)^2
\int d^2z_1|z_1|^{2\D_\infty -2\D_0 -2}\nonumber\\
&\times& \int d^2u|u|\langle f|T\left( \cL_{int}(1,1)
\cL_{int}(u,\bar u)\right) |i\rangle .
\nonumber\eea
The integral over $z_1$ is obviously divergent. To understand the meaning of
this divergency one should remember that we made the Wick rotation. Coming back
to the $\s ,\tau$-coordinates on the cylinder, we get for the integral over
$z_1$
$$
\int d^2z_1|z_1|^{2\D_\infty -2\D_0 -2}\to
i\int d\tau d\s \e^{2i\tau (\D_\infty -\D_0 )}.
$$
Integration over $\s$ and $\tau$ gives us the conservation law for the
light-cone momenta $k_i^-$
$$
\int d\tau d\s \e^{2i\tau (\D_\infty -\D_0 )}=
4N(2\pi )^2\d (k_1^-+k_2^-+k_3^-+k_4^-).$$
Thus, the S-matrix element is equal to
\bea
\langle f|S|i\rangle =-i2\lambda^2
N^3\d (k_1^-+k_2^-+k_3^-+k_4^-)
\int d^2u|u|\langle f|T\left( \cL_{int}(1,1)
\cL_{int}(u,\bar u)\right) |i\rangle .
\label{matel3}
\eea
So, to find the S-matrix element one has to calculate the correlation function
\bea
&&F(u,\bar u)=\langle f|T\left( \cL_{int}(1,1)
\cL_{int}(u,\bar u)\right) |i\rangle \nonumber\\
&&=C_0C_\infty \sum_{I<J;K<L}\langle
V_{[g_\infty ]}[\k_3,\z_3;\k_4,\z_4](\infty )T\left(
V_{IJ}(1,1)V_{KL}(u,\bar u)\right)
V_{[g_0]}[\k_1,\z_1;\k_2,\z_2](0,0)\rangle .
\la{fuu}
\eea
In what follows we assume for definiteness that $n_0<n_\infty$ and $|u|<1$.

By using the definition (\ref{invver}) of $V_{[g]}$, and taking
into account that the interaction vertex is $S_N$-invariant, and that any
correlation function of vertex operators is invariant with respect to the global action
of the symmetric group
\be
\langle
V_{g_1}V_{g_2}\cdots V_{g_n}\rangle =
\langle V_{h^{-1}g_1h}V_{h^{-1}g_2h}\cdots V_{h^{-1}g_nh}\rangle ,
\label{invcor}
\ee
we rewrite the correlation function in the form
\bea 
F(u,\bar u) =\frac {C_0C_\infty }{N!}\sum_{h_\infty\in S_N}
\sum_{I<J;K<L} \langle V_{h_\infty ^{-1}g_\infty
h_\infty}[\k_3,\z_3;\k_4,\z_4](\infty ) V_{IJ}(1,1)V_{KL}(u,\bar
u)V_{g_0}[\k_1,\z_1;\k_2,\z_2](0,0)\rangle .
\la{fuu1}
\eea
Let us note that the correlation function
\be
\langle V_{g_1}(\infty ) V_{g_2}(1,1)V_{g_3}(u,\bar u)V_{g_4}(0,0)\rangle
\la{corr}
\ee
does not vanish only if
\be
g_1g_2g_3g_4=1\quad \mbox{or}\quad g_1g_4g_3g_2=1.
\la{aa}
\ee
It can be seen as follows. Due to the OPE (\ref{OPE}) of $V_g$, in the limit
$u\to 0$ the correlation function (\ref{corr}) reduces to the sum of three-point
correlation functions
$\langle V_{g_1}V_{g_2}V_{g_3g_4}\rangle $ and
$\langle V_{g_1}V_{g_2}V_{g_4g_3}\rangle $. This sum does not vanish
if one of the following equations is fulfilled:
\be
g_1g_2g_3g_4=1,\quad g_1g_3g_4g_2=1,\quad  g_1g_2g_4g_3=1,\quad g_1g_4g_3g_2=1.
\la{aa1}
\ee
From the other side in the limit $u\to 1$ one gets the sum of the correlation functions
$\langle V_{g_1}V_{g_2g_3}V_{g_4}\rangle $ and
$\langle V_{g_1}V_{g_3g_2}V_{g_4}\rangle $. This sum does not vanish
if \be g_1g_2g_3g_4=1,\quad g_1g_4g_2g_3=1,\quad
g_1g_3g_2g_4=1,\quad g_1g_4g_3g_2=1.  \la{aa2} \ee
Comparing
eqs.(\ref{aa1}) and (\ref{aa2}), one obtains eq.(\ref{aa}).

Thus, every summand in (\ref{fuu1}) is not equal to zero in the following
two cases:
$$
h_\infty ^{-1}g_\infty h_\infty g_0g_{KL}g_{IJ}=1,\qquad
h_\infty ^{-1}g_\infty h_\infty g_{IJ}g_{KL}g_0=1.
$$
On the other hand, one can
express the correlation functions with the second monodromy condition
via the correlation functions with the first one.
Indeed, since the action and the interaction vertex
of the model is invariant under the world-sheet parity transformation
$z\to \bar{z}$ and the space-reflection $X^3\to -X^3$, and
the graviton vertex operators
$V_g[\{\k_{\al},\z_{\al}\}]$ transform into
$\tilde{V}_{g^{-1}}[\{\k_{\al},\z_{\al}\}]\equiv
V_{g^{-1}}[\{\tilde{\k}_{\al},\tilde{\z}_{\al}\}]$
their correlation functions satisfy the following equality:
\bea
\langle V_{g_{IJ}g_{KL}g_0^{-1}}V_{IJ}V_{KL}V_{g_0}\rangle =
\langle
\tilde{V}_{g_0g_{KL}g_{IJ}}V_{IJ}V_{KL}\tilde{V}_{g_0^{-1}}\rangle .
\nonumber
\eea
Now taking into
account eqs.(\ref{invcor}) and that the elements $g$ and
$g^{-1}$ belong to the same conjugacy class, one obtains
\bea
\langle V_{g_{IJ}g_{KL}g_0^{-1}}V_{IJ}V_{KL}V_{g_0}\rangle =
\langle \tilde{V}_{g_0^{-1}g_{K'L'}g_{I'J'}}V_{I'J'}V_{K'L'}\tilde{V}_{g_0}\rangle ,
\nonumber
\eea
where $g_{I'J'}=hg_{IJ}h^{-1}$, $g_{K'L'}=hg_{KL}h^{-1}$, and
the element $h$ is such that $g_0^{-1}=h^{-1}g_0h$.
It is now clear that the contribution of the terms satisfying
the second monodromy condition coincides with the one of the terms
satisfying the first monodromy condition after the replacement
$\k_{\al}\to \tilde{\k}_{\al}$ and $\z_{\al}\to \tilde{\z}_{\al}$.
It is obvious that due to the $SO(8)$ invariance the correlation function
(\ref{fuu}) can depend only on the scalar products of momenta
and polarizations of the gravitons and, therefore, is invariant
under space reflections. Thus, the contributions of the terms
satisfying the first and the second monodromy conditions coincide.

Schematically, the function $F(u,\bar u)$ given by a sum of correlation functions of twist
fields can be  represented as
\bea
{\cal S}=\sum_{h_\infty\in S_N}\sum_{I<J;K<L}
\langle V_{h_\infty^{-1} g_\infty h_\infty}
V_{IJ}V_{KL}V_{g_0}\rangle ,
\nonumber\eea
where the elements $h_\infty ,g_{IJ},g_{KL}$ solve the equation
$h_\infty ^{-1}g_\infty h_\infty g_{IJ}g_{KL}g_0=1$.

\noindent We can fix the values of the indices $K$ and $L$ by using the action
of the centralizer of $g_0$ and the invariance (\ref{invcor}) of the correlation functions
\bea
{\cal S}=\sum_{h_\infty\in S_N}\sum_{I<J}&&\left(
n_0(N-n_0)\langle V_{h_\infty^{-1} g_\infty h_\infty}
V_{IJ}V_{n_0N} V_{g_0}\rangle \right.\nonumber\\
&&+\left. (N-n_0)\langle V_{h_\infty^{-1} g_\infty h_\infty}
V_{IJ}V_{n_\infty N}V_{g_0}\rangle \right.\nonumber\\
&&+\left. (N-n_0)\langle V_{h_\infty^{-1} g_\infty h_\infty}
V_{IJ}V_{n_0+n_\infty ,N}V_{g_0}\rangle \right) .
\la{s1}
\eea
The first term in (\ref{s1}) corresponds to the joining of two incoming strings
and the factor $n_0(N-n_0)$ appears since in this case the index $K$ can take
$n_0$ values, $K=1,...,n_0$, and the index $L$ takes $N-n_0$ values,
$L=n_0+1,...,N$. To fix $K=n_0$ and $L=N$ we have to use all elements of
$C_{g_0}$. The second and the third terms correspond to the splitting of the
string of length $N-n_0$ into two strings of lengths $n_\infty -n_0$ and
$N-n_\infty$, and $N-n_0-n_\infty$ and $n_\infty$ respectively. In these cases
to fix the values of $K$ and $L$ one should use $N-n_0$ elements of the
subgroup $\Z_{N-n_0}$ of $C_{g_0}$ that does not act on the cycle $(n_0)$.

Eq.(\ref{s1}) can be further rewritten in the form
\bea
{\cal S}&=&n_0(N-n_0)n_\infty (N-n_\infty )
\left(\sum_{I=1}^{n_\infty}
\langle V_{g_\infty (I)}
V_{I,I+N-n_\infty}V_{n_0N}V_{g_0}\rangle \right.\nonumber\\
&+&\sum_{I=1}^{N-n_\infty}\langle V_{g_\infty (I)}
V_{I,I+n_\infty}V_{n_0 N}V_{g_0}\rangle +
\sum_{J=n_0+1}^{n_\infty}
\langle V_{g_\infty (J)}
V_{n_0J}V_{n_\infty N}V_{g_0}\rangle
\nonumber\\
&+&\left.\sum_{J=n_0+n_\infty +1}^{N}
\langle V_{g_\infty (J)}
V_{n_0J}V_{n_0+n_\infty ,N}V_{g_0}
\rangle \right),
\la{s2}
\eea
where the elements $g_\infty $ have to be found from the equation $g_\infty
g_{IJ}g_{KL}g_0=1$.
The  diagramms corresponding to these four terms are depicted in Fig.1.
\begin{figure}[t]
\special{em:linewidth 0.4pt}
\unitlength 1.00mm
\linethickness{0.4pt}
\begin{picture}(130.33,145.62)(-10,0)
\put(13.67,145.00){\circle{13.23}}
\put(13.67,132.00){\circle{4.06}}
\put(140.67,132.00){\circle{6.32}}
\put(140.67,145.00){\circle{8.67}}
\put(13.67,105.00){\circle{12.81}}
\put(13.67,90.00){\circle{4.81}}
\put(13.33,65.00){\circle{11.66}}
\put(13.67,50.00){\circle{4.81}}
\put(13.67,25.00){\circle{11.33}}
\put(13.67,10.00){\circle{4.85}}
\put(141.00,90.00){\circle{8.03}}
\put(141.33,105.00){\circle{4.71}}
\put(141.33,65.00){\circle{8.67}}
\put(141.33,50.00){\circle{6.04}}
\put(142.00,10.00){\circle{8.00}}
\put(142.00,25.00){\circle{6.00}}
\emline{15.00}{151.33}{1}{38.92}{149.91}{2}
\emline{38.92}{149.91}{3}{62.76}{148.95}{4}
\emline{62.76}{148.95}{5}{86.54}{148.45}{6}
\emline{86.54}{148.45}{7}{114.18}{148.44}{8}
\emline{114.18}{148.44}{9}{140.00}{149.00}{10}
\emline{14.67}{138.33}{11}{30.66}{138.19}{12}
\emline{30.66}{138.19}{13}{43.17}{137.85}{14}
\emline{43.17}{137.85}{15}{54.94}{137.32}{16}
\emline{54.94}{137.32}{17}{60.33}{137.00}{18}
\emline{14.33}{133.67}{19}{59.67}{137.00}{20}
\emline{14.33}{130.33}{21}{40.03}{131.10}{22}
\emline{40.03}{131.10}{23}{64.87}{131.40}{24}
\emline{64.87}{131.40}{25}{84.91}{131.28}{26}
\emline{84.91}{131.28}{27}{104.36}{130.82}{28}
\emline{104.36}{130.82}{29}{123.21}{130.04}{30}
\emline{123.21}{130.04}{31}{140.33}{129.00}{32}
\emline{140.33}{140.67}{33}{90.67}{139.33}{34}
\emline{140.33}{134.67}{35}{131.48}{135.18}{36}
\emline{131.48}{135.18}{37}{119.64}{136.08}{38}
\emline{119.64}{136.08}{39}{106.93}{137.29}{40}
\emline{106.93}{137.29}{41}{91.67}{139.00}{42}
\emline{15.67}{111.00}{43}{42.23}{109.33}{44}
\emline{42.23}{109.33}{45}{69.46}{108.15}{46}
\emline{69.46}{108.15}{47}{97.39}{107.45}{48}
\emline{97.39}{107.45}{49}{130.13}{107.25}{50}
\emline{130.13}{107.25}{51}{141.33}{107.33}{52}
\emline{14.00}{87.67}{53}{49.05}{88.08}{54}
\emline{49.05}{88.08}{55}{84.79}{87.79}{56}
\emline{84.79}{87.79}{57}{121.23}{86.81}{58}
\emline{121.23}{86.81}{59}{141.00}{86.00}{60}
\emline{14.00}{98.67}{61}{60.00}{96.00}{62}
\emline{14.00}{92.33}{63}{19.96}{92.62}{64}
\emline{19.96}{92.62}{65}{28.84}{93.20}{66}
\emline{28.84}{93.20}{67}{44.12}{94.51}{68}
\emline{44.12}{94.51}{69}{58.67}{96.00}{70}
\emline{141.67}{102.67}{71}{90.00}{99.67}{72}
\emline{91.67}{99.67}{73}{93.51}{99.17}{74}
\emline{93.51}{99.17}{75}{95.41}{98.69}{76}
\emline{95.41}{98.69}{77}{97.39}{98.23}{78}
\emline{97.39}{98.23}{79}{99.43}{97.80}{80}
\emline{99.43}{97.80}{81}{101.55}{97.38}{82}
\emline{101.55}{97.38}{83}{103.73}{96.98}{84}
\emline{103.73}{96.98}{85}{105.99}{96.61}{86}
\emline{105.99}{96.61}{87}{108.31}{96.25}{88}
\emline{108.31}{96.25}{89}{110.70}{95.92}{90}
\emline{110.70}{95.92}{91}{113.16}{95.61}{92}
\emline{113.16}{95.61}{93}{115.69}{95.32}{94}
\emline{115.69}{95.32}{95}{118.29}{95.05}{96}
\emline{118.29}{95.05}{97}{123.70}{94.57}{98}
\emline{123.70}{94.57}{99}{129.39}{94.17}{100}
\emline{129.39}{94.17}{101}{135.36}{93.86}{102}
\emline{135.36}{93.86}{103}{140.33}{93.67}{104}
\emline{14.00}{70.33}{105}{24.92}{69.41}{106}
\emline{24.92}{69.41}{107}{36.06}{68.65}{108}
\emline{36.06}{68.65}{109}{47.43}{68.06}{110}
\emline{47.43}{68.06}{111}{62.92}{67.53}{112}
\emline{62.92}{67.53}{113}{78.80}{67.30}{114}
\emline{78.80}{67.30}{115}{95.07}{67.36}{116}
\emline{95.07}{67.36}{117}{111.73}{67.71}{118}
\emline{111.73}{67.71}{119}{128.78}{68.37}{120}
\emline{128.78}{68.37}{121}{141.00}{69.00}{122}
\emline{13.67}{47.67}{123}{34.06}{48.62}{124}
\emline{34.06}{48.62}{125}{54.32}{49.18}{126}
\emline{54.32}{49.18}{127}{74.45}{49.35}{128}
\emline{74.45}{49.35}{129}{94.45}{49.12}{130}
\emline{94.45}{49.12}{131}{114.32}{48.50}{132}
\emline{114.32}{48.50}{133}{130.12}{47.73}{134}
\emline{130.12}{47.73}{135}{141.67}{47.00}{136}
\emline{13.67}{59.33}{137}{69.51}{56.12}{138}
\emline{69.51}{56.12}{139}{89.67}{55.33}{140}
\emline{13.67}{52.33}{141}{88.67}{55.33}{142}
\emline{141.00}{52.67}{143}{107.00}{56.24}{144}
\emline{107.00}{56.24}{145}{77.03}{59.96}{146}
\emline{77.03}{59.96}{147}{67.00}{61.33}{148}
\emline{68.00}{61.00}{149}{118.93}{60.56}{150}
\emline{118.93}{60.56}{151}{141.33}{60.67}{152}
\emline{13.67}{30.67}{153}{32.06}{29.38}{154}
\emline{32.06}{29.38}{155}{51.05}{28.41}{156}
\emline{51.05}{28.41}{157}{70.64}{27.76}{158}
\emline{70.64}{27.76}{159}{94.94}{27.41}{160}
\emline{94.94}{27.41}{161}{120.10}{27.53}{162}
\emline{120.10}{27.53}{163}{142.00}{28.00}{164}
\emline{13.67}{7.67}{165}{23.79}{8.22}{166}
\emline{23.79}{8.22}{167}{37.89}{8.73}{168}
\emline{37.89}{8.73}{169}{52.70}{8.97}{170}
\emline{52.70}{8.97}{171}{68.22}{8.96}{172}
\emline{68.22}{8.96}{173}{88.61}{8.58}{174}
\emline{88.61}{8.58}{175}{114.53}{7.59}{176}
\emline{114.53}{7.59}{177}{142.33}{6.00}{178}
\emline{14.00}{19.33}{179}{85.00}{15.33}{180}
\emline{14.00}{12.33}{181}{41.68}{13.81}{182}
\emline{41.68}{13.81}{183}{84.00}{15.33}{184}
\emline{67.00}{21.67}{185}{97.00}{18.07}{186}
\emline{97.00}{18.07}{187}{131.15}{14.63}{188}
\emline{131.15}{14.63}{189}{142.00}{13.67}{190}
\emline{67.33}{21.67}{191}{119.89}{21.69}{192}
\emline{119.89}{21.69}{193}{142.00}{22.00}{194}
\put(146.00,132.00){\makebox(0,0)[lc]{$n_{\infty},{\bf k}_3$}}
\put(146.33,145.00){\makebox(0,0)[lc]{$N-n_{\infty},{\bf k}_4$}}
\put(5.33,145.00){\makebox(0,0)[rc]{$N-n_0,{\bf k}_2$}}
\put(9.67,132.00){\makebox(0,0)[rc]{$n_0,{\bf k}_1$}}
\put(5.33,145.00){\makebox(0,0)[rc]{$N-n_0,{\bf k}_2$}}
\put(6.00,105.00){\makebox(0,0)[rc]{$N-n_0,{\bf k}_2$}}
\put(10.33,90.00){\makebox(0,0)[rc]{$n_0,{\bf k}_1$}}
\put(6.33,64.67){\makebox(0,0)[rc]{$N-n_0,{\bf k}_2$}}
\put(10.00,49.67){\makebox(0,0)[rb]{$n_0,{\bf k}_1$}}
\put(7.00,24.67){\makebox(0,0)[rc]{$N-n_0,{\bf k}_2$}}
\put(10.33,9.67){\makebox(0,0)[rc]{$n_0,{\bf k}_1$}}
\put(145.00,104.67){\makebox(0,0)[lc]{$n_{\infty},{\bf k}_3$}}
\put(146.67,90.00){\makebox(0,0)[lc]{$N-n_{\infty},{\bf k}_4$}}
\put(147.00,65.00){\makebox(0,0)[lc]{$N-n_{\infty},{\bf k}_4$}}
\put(145.67,50.00){\makebox(0,0)[lc]{$n_{\infty},{\bf k}_3$}}
\put(146.67,25.00){\makebox(0,0)[lc]{$n_{\infty},{\bf k}_3$}}
\put(147.33,10.00){\makebox(0,0)[lc]{$N-n{\infty},{\bf k}_4$}}
\end{picture}
\caption{The diagramm representation of different correlation functions in
eq.(4.32)}
\label{fig1}
\end{figure}
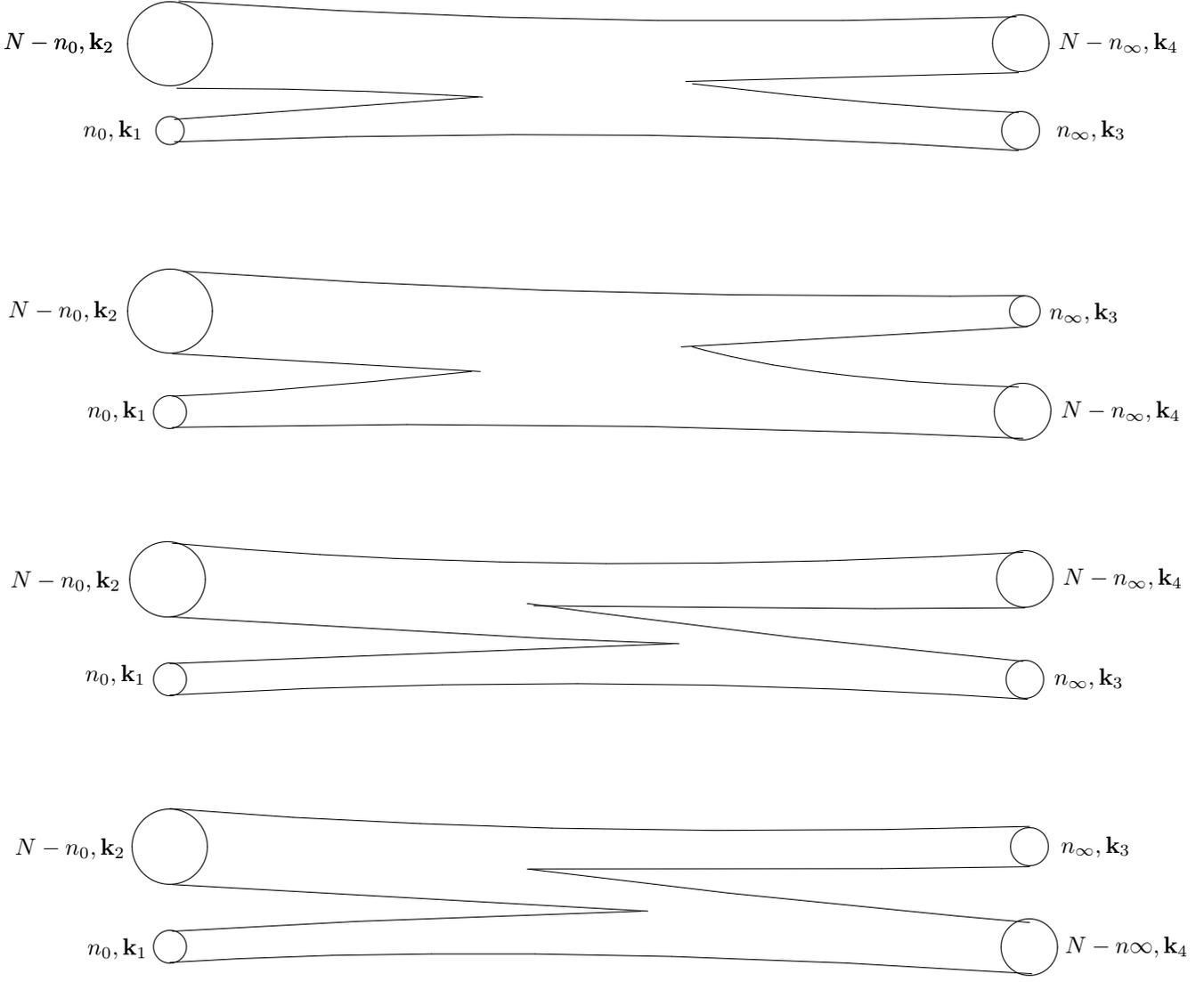

So, we need to compute the correlation functions (and the same correlation functions with the
interchange $u\leftrightarrow 1$)
\be
G(u,\bar u)=
\langle V_{ g_\infty}[\k_3,\z_3;\k_4,\z_4](\infty )
V_{IJ}(1,1)V_{KL}(u,\bar u) V_{g_0}[\k_1,\z_1,\k_2,\z_2](0,0)\rangle ,
\la{guu}
\ee
where all possible elements $g_\infty ,g_{IJ},g_{KL},g_0$ are listed in
eq.(\ref{s2}).

To calculate the correlation function (\ref{guu}) we employ the stress-energy tensor
method \cite{DFMS}. The idea of the method is as follows. Suppose that one
knows the following ratio
\bea
f(z,u)=
\frac {\langle T(z)\phi_\infty (\infty )\phi_1(1)\phi_2(u)\phi_0(0)\rangle }
{\langle \phi_\infty (\infty )\phi_1(1)\phi_2(u)\phi_0(0)\rangle },
\nonumber
\eea
where $T(z)$ is the stress-energy tensor and $\phi$ are primary fields. Taking
into account that the OPE of $T(z)$ with any primary field has the form
\bea
T(z)\phi (0)=\frac {\D}{z^2}\phi (0) +\frac {1}{z}\partial\phi (0)+\cdots ,
\nonumber
\eea
one gets a differential equation on the correlation function
$G(u,\bar u)=\langle \phi_\infty (\infty )\phi_1(1)\phi_2(u)\phi_0(0)\rangle$
\bea
\partial _u\log G(u,\bar u)=H(u,\bar u),
\nonumber\eea
where $H(u,\bar u)$ is the second term in the decomposition of the function
$f(z,u)$ in the vicinity of $u$
\bea
f(z,u)=\frac {\D_2}{(z-u)^2} +\frac {1}{z-u}H(u,\bar u)+\cdots .
\nonumber
\eea
In the same way one gets the second equation on $G(u,\bar u)$ by using
the stress-energy tensor $\bar T(\bar z)$
\bea
\partial _{\bar u}\log G(u,\bar u)=\bar H(u,\bar u).
\nonumber\eea
A solution of these two equations determines the correlation function
$G(u,\bar u)$ up to a constant.

It turns out that for the correlation functions we consider the functions
$H(u,\bar u)$ and $\bar H(u,\bar u)$ are holomorphic
and antiholomorphic functions respectively. Therefore, the correlation function
$G(u,\bar u)$ admits a factorization $G(u,\bar u)=G(u)\bar{G}(\bar u)$.
Moreover, since the stress-energy tensor is a sum of the bosonic
and fermionic ones, the function $G(u)$ admits further
factorization $G(u)=G_b(u)G_f(u)$. Thus,
the correlation function $G(u,\bar u)$ acquires the form
\be
\la{decom}
G(u,\bar u)=G_b(u)G_f(u)\bar{G}_b(\bar u)\bar{G}_f(\bar u),
\ee
where, e.g., $G_b(u)$ is a contribution of the bosonic left-moving sector
to the correlation function. By using the stress-energy tensor method one can
compute each multiplier on the r.h.s. of (\ref{decom}) up to a constant.
Since the vertex operators can not be represented as a tensor product
of bosonic and fermionic twist fields,
only the overall normalization constant for $G(u,\bar u)$
can be found exactly. As to the individual constants, below we show
that they can be determined only up to phases.

In the next two sections we calculate the contribution of the
bosonic and fermionic correlation functions occuring in (\ref{decom}).

\section{Bosonic correlation functions}
\setcounter{equation}{0}
By using the definition of the vertex operators one can see that
the contribution of the bosonic left-moving sector to correlation function
(\ref{guu}) is given by
\bea
\nonumber
G_b^{ij}(u)=\l \s_{g_{\infty}}[\k_3/2,\k_4/2](\infty)
\t^{i}_{IJ}(1)\t^{j}_{KL}(u)\s_{g_0}[\k_1/2,\k_2/2](0) \r .
\eea
According to the definition (\ref{tau}) of $\t$  this correlation function can
be written as the following limit
\be
\la{Gbu1}
G_b^{ij}(u)=\lim_{z\to 1,w\to u}(z-1)^{1/2}(w-u)^{1/2} \l \partial
X^i_I(z)\partial X^j_K(w) \s_{g_{\infty}}[\k_3/2,\k_4/2](\infty)
\s_{IJ}(1)\s_{KL}(u)\s_{g_0}[\k_1/2,\k_2/2](0) \r .
\ee
Thus,  the calculation of $G_b^{ij}$ is reduced to calculation
of the Green function
\bea
G^{ij}_{MS}(z,w)
&=&
\frac {\langle \partial X^i_M(z)\partial X^j_S(w)\s_{
g_\infty}[\k_3/2,\k_4/2](\infty ) \s_{IJ}(1)\s_{KL}(u)
\s_{g_0}[\k_1/2,\k_2/2](0)\rangle }
{\langle \s_{g_\infty}[\k_3/2,\k_4/2](\infty )
\s_{IJ}(1)\s_{KL}(u) \s_{g_0}[\k_1/2,\k_2/2](0)\rangle }
\nonumber\\ &\equiv&\langle\langle
\partial X^i_M(z)\partial X^j_S(w)\rangle\rangle  .
\nonumber\eea
and the correlation function
\bea
\la{cor}
G_b(u)=
\langle \s_{g_\infty}[\k_3/2,\k_4/2](\infty )
\s_{IJ}(1)\s_{KL}(u) \s_{g_0}[\k_1/2,\k_2/2](0)\rangle .
\eea

We start with considering the more general correlation function
\bea
G(u)=
\langle \s_{g_\infty}[\p_3,\p_4](\infty )
\s[\p_5]_{IJ}(1)\s[\p_6]_{KL}(u) \s_{g_0}[\p_1,\p_2](0)\rangle ,
\la{gencor}
\eea
and the corresponding Green function
\bea
G^{ij}_{MS}(z,w)
&=&
\frac {\langle \partial X^i_M(z)\partial X^j_S(w)
\s_{g_\infty}[\p_3,\p_4](\infty ) \s[\p_5]_{IJ}(1)\s[\p_6]_{KL}(u)
\s_{g_0}[\p_1,\p_2](0)\rangle }
{\langle \s_{g_\infty}[\p_3,\p_4](\infty )
\s[\p_5]_{IJ}(1)\s[\p_6]_{KL}(u) \s_{g_0}[\p_1,\p_2](0)\rangle },
\nonumber
\eea
for $D$ bosonic fields and arbitrary momenta $\p_5$ and $\p_6$
keeping in mind application to calculation of the fermionic
correlation functions.

These Green functions have non-trivial monodromies around points
$\infty ,1,u$ and $0$, and, in fact, are different branches of
one multi-valued function. However, this function is single-valued on the
sphere that is obtained by gluing the fields $X^i_I$ at $z=0$ and $z=\infty$.
Thus to construct $G^{ij}_{MS}(z,w)$ we introduce the following
map from this sphere onto the original one:
\be
z=\left(\frac {t}{t_1}\right)^{n_0}
\left(\frac{t-t_0}{t_1-t_0}\right)^{N-n_0}
\left(\frac {t_1-t_\infty}{t-t_\infty}\right)^{N-n_{\infty}} \equiv u(t).
\la{map}
\ee
Here the
points $t=0$ and $t=t_0$ are mapped  to the point $z=0$; $t=\infty $,
$t=t_\infty\to z=\infty$, $t=t_1\to z=1$ and $t=x\to z=u$ (see
Fig.2).
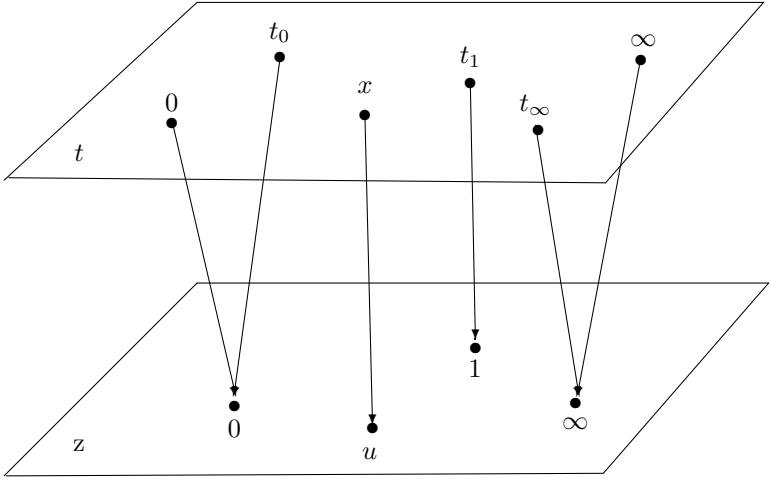
\begin{figure} 
\special{em:linewidth 0.4pt}
\unitlength 1.00mm
\linethickness{0.4pt}
\begin{picture}(128.67,67.67)(0,30)
\emline{27.00}{34.67}{1}{52.67}{60.33}{2}
\emline{52.67}{60.33}{3}{128.67}{60.33}{4}
\emline{128.67}{60.33}{5}{106.67}{35.00}{6}
\emline{106.67}{35.00}{7}{27.33}{34.67}{8}
\emline{27.00}{74.00}{9}{52.67}{97.67}{10}
\emline{52.67}{97.67}{11}{128.00}{97.67}{12}
\emline{128.00}{97.67}{13}{107.00}{73.67}{14}
\emline{107.00}{73.67}{15}{27.67}{74.33}{16}
\put(57.67,46.00){\vector(1,-4){0.2}}
\emline{49.33}{82.00}{17}{57.67}{46.00}{18}
\put(103.33,46.00){\vector(1,-4){0.2}}
\emline{97.67}{81.33}{19}{103.33}{46.00}{20}
\put(103.33,46.00){\vector(-1,-4){0.2}}
\emline{111.67}{90.33}{21}{103.33}{46.00}{22}
\put(76.00,41.67){\vector(0,-1){0.2}}
\emline{75.00}{83.00}{23}{76.00}{41.67}{24}
\put(89.67,53.00){\vector(0,-1){0.2}}
\emline{89.00}{87.33}{25}{89.67}{53.00}{26}
\put(75.00,86.33){\makebox(0,0)[cc]{$x$}}
\put(89.00,90.33){\makebox(0,0)[cc]{$t_1$}}
\put(97.67,84.00){\makebox(0,0)[cc]{$t_{\infty}$}}
\put(112.00,92.67){\makebox(0,0)[cc]{$\infty$}}
\put(37.00,77.67){\makebox(0,0)[cc]{$t$}}
\put(37.00,38.67){\makebox(0,0)[cc]{z}}
\put(49.33,84.33){\makebox(0,0)[cc]{$0$}}
\put(57.67,41.00){\makebox(0,0)[cc]{$0$}}
\put(75.67,37.67){\makebox(0,0)[cc]{$u$}}
\put(103.00,41.67){\makebox(0,0)[cc]{$\infty$}}
\put(49.67,81.67){\rule{0.00\unitlength}{0.00\unitlength}}
\put(63.33,90.33){\rule{0.00\unitlength}{0.00\unitlength}}
\put(75.33,82.67){\rule{0.00\unitlength}{0.00\unitlength}}
\put(88.67,87.33){\rule{0.00\unitlength}{0.00\unitlength}}
\put(97.67,81.00){\rule{0.00\unitlength}{0.00\unitlength}}
\put(111.67,90.00){\rule{0.00\unitlength}{0.00\unitlength}}
\put(63.67,93.67){\makebox(0,0)[cc]{$t_0$}}
\put(89.67,49.00){\makebox(0,0)[cc]{$1$}}
\put(49.33,81.67){\circle*{0.67}}
\put(63.67,90.33){\circle*{1.49}}
\put(49.33,81.67){\circle*{1.49}}
\put(75.00,82.67){\circle*{1.33}}
\put(89.00,87.00){\circle*{0.00}}
\put(89.00,87.00){\circle*{1.33}}
\put(98.00,80.67){\circle*{1.33}}
\put(111.67,90.00){\circle*{1.49}}
\put(57.67,46.00){\vector(-1,-4){0.2}}
\emline{63.67}{90.00}{27}{57.67}{46.00}{28}
\put(57.67,44.00){\circle*{1.33}}
\put(76.00,41.00){\circle*{1.49}}
\put(89.67,51.67){\circle*{1.33}}
\put(103.00,44.33){\circle*{1.49}}
\end{picture}
\label{fig2}
\caption{The $N$-fold covering of the $z$-sphere by the $t$-sphere.}
\end{figure}
In what follows we often use the notation $\Omega_A$ to refer to the
set of the branch points:
$\Omega_1=0,~\Omega_2=t_0$, $\Omega_3=\infty$,
$\Omega_4=t_\infty,~\Omega_5=t_1$ and $\Omega_6=x$.

The map (\ref{map}) may be viewed as the
$N$-fold covering of the $z$-plain by the $t$-sphere on which the
Green function  is single-valued. The more detailed discussion of
eq.(\ref{map}) is presented in the Appendix B.

Due to the projective transformations, the positions of
the points $t_0,t_\infty ,t_1$ depend on $x$ and it is convinient to choose
this dependence as follows
\bea
t_0&=&x-1,\nonumber\\
t_\infty &=&x-\frac {(N-n_\infty )x}{(N-n_0)x+n_0},\nonumber\\
t_1&=&\frac {N-n_0-n_\infty }{n_\infty}+\frac {n_0x}{n_\infty}-
\frac {N(N-n_\infty )x}{n_\infty ((N-n_0)x+n_0)}.
\nonumber\eea
This choice leads to the following expression for
the rational function $u(x)$
\bea
u=u(x)&=&(n_0-n_{\infty})^{n_0-n_{\infty}}\frac{n_{\infty}^{n_{\infty}}}
{n_0^{n_0}}\left(\frac{N-n_0}{N-n_{\infty}}\right)^{N-n_{\infty}}
\left(\frac{x+\frac{n_0}{N-n_0}}{x-1}\right)^N\nonumber\\
&\times&\left(\frac{x-\frac{N-n_0-n_{\infty}}{N-n_0}}{x}\right)^{N-n_0-n_{\infty}}
\left(x-\frac{n_0}{n_0-n_{\infty}}\right)^{n_0-n_{\infty}}.
\la{ux}
\eea
Since $n_0<n_{\infty}$, the map $u(x)$ can be treated as the
$2(N-n_0)$-fold covering of the $u$-sphere by the $x$-sphere, that
means that an equation $u(x)=u$ has $2(N-n_0)$ different
solutions. It is worthwhile to note that
this number coincides with the number of nontrivial correlation functions in eq.(\ref{s2})
and, therefore different roots of eq.(\ref{ux}) correspond to different
correlation functions (\ref{s2}). We see that the $t$-sphere can be represented as the
union of $2(N-n_0)$ domains, and each domain $V_{IJKL}$ contains the points $x$
corresponding to the correlation function (\ref{guu}). If we take on the
$u$-plain the appropriate system of cuts, then every root of eq.(\ref{ux})
realizes a one-to-one conformal mapping of the cut $u$-plain onto
the corresponding domain $V_{IJKL}$.

Let us now choose some root of eq.(\ref{ux}).
One can always cut the $z$-sphere and numerate
the roots $t_R(z)$ of eq.(\ref{map}) in such a way that they have the
same monodromies as the fields $X$ do.
Then the Green functions are
obviously not equal to zero only if
the momentum conservation law
$\p_1+\p_2+\p_3+\p_4+\p_5+\p_6=0$ is fulfilled,
and are given by
\bea
\la{green1}
G_{MS}^{ij}(z,w)=-\d^{ij}\frac{t_M'(z)t_S'(w)}{(t_M(z)-t_S(w))^{2}}
-\sum_{AB}\frac{p_A^ip_B^j t_M'(z)t_S'(w)}
{(t_M(z)-\Omega_A)(t_S(w)-\Omega_B)}.
\eea
Since $\Omega_3=\infty$ the corresponding terms
in the sum (\ref{green1}) are absent.

One can easily check that these functions have the singularity
$-\frac {\d^{ij}\d_{MS}}{(z-w)^{2}}$ in the vicinity $z-w=0$ and proper
monodromies around the points $z=\infty ,1,u,0$.

Recall that the stress-energy tensor is defined as
\bea
T(z)= -\frac 12 \lim _{w\to z}\sum_{i=1}^{D}\sum_{I=1}^N\left(
\partial X_I^i(z)\partial X_I^i(w) +\frac {1}{(z-w)^2}\right).
\nonumber
\eea
By using this definition and eq.(\ref{green1}), one gets
\footnote{If all $\p_i=0$, the expectation value of
$T(z)$ in the presence of twist fields can be equivalently found by
mapping with $t_M(z)$ the stress-energy tensor on the $t$-sphere onto
the $z$-sphere with the subsequent summation over $M$ (see
e.g.\cite{DFMS})}
\bea
\langle\langle T(z)\rangle\rangle =
\sum_M
\frac {D}{12}\left(\left(\frac {t_M''(z)}{t_M'(z)}\right)'-
\frac {1}{2}\left(\frac {t_M''(z)}{t_M'(z)}\right)^2\right)
+\sum_{AB,M}\frac{\p_A \p_B
(t_M'(z))^2}
{2(t_M(z)-\Omega_A)(t_M(z)-\Omega_B)}.
\nonumber
\eea

The term
$$\left(\frac {t''}{t'}\right) '-
\frac {1}{2}\left(\frac {t''}{t'}\right)^2=
\frac {t'''}{t'}-\frac {3}{2}\left(\frac {t''}{t'}\right)^2$$
is the Schwartz derivative as one could expect from the very beginning. To get
the differential equation on the correlation function (\ref{gencor}) 
one should expand
$\langle\langle T(z)\rangle\rangle $ in the vicinity of $z=u$. This expansion
is given by
\bea
\la{dect}
&&
\l\l T(z)\r\r=\frac{\p_6^2+D/4}{4}\frac{1}{(z-u)^2}
+\frac{1}{u(z-u)}\left(-\frac{\p_6^2+D/4}{4}+
\frac{9\p_6^2+3D/2}{16}\frac{a_1^2}{a_0^3}
-\frac{\p_6^2+D/4}{2}\frac{a_2}{a_0^2}
\right.\\
\nonumber
&&
\left.
+\sum_{A,B=1}^{5}\frac{\p_A\p_B}{4a_0(x-\Omega_A)(x-\Omega_B)}
-\sum_{A=1}^{5}\frac{\p_6\p_A}{2a_0(x-\Omega_A)^2}
\left(1+\frac{3a_1(x-\Omega_A)}{a_0}\right) \right)+\cdots .
\eea

Here the coefficients $a_k$ are defined as follows
\bea
a_k=\frac {(-1)^{k-1}}{k+2}\left( \frac {n_0}{x^{k+2}}+
\frac {N-n_0}{(x-t_0)^{k+2}}-\frac {N-n_\infty}{(x-t_\infty)^{k+2}}\right) .
\nonumber
\eea
The first term shows that the conformal dimension of the twist field
$\s_{KL}[\p_6]$ is equal to
$\frac {D}{16}+\frac {\p_6^2}{4}$, as it should be, and the other
terms lead to the following differential equation on $G(u)$
\bea
\nonumber
u\partial_u \log G(u)&=&
-\frac{\p_6^2+D/4}{4}+
\frac{9\p_6^2+3D/2}{16}\frac{a_1^2}{a_0^3}
-\frac{\p_6^2+D/4}{2}\frac{a_2}{a_0^2}
\\
&&
+\sum_{A,B=1}^{5}\frac{\p_A\p_B}{4a_0(x-\Omega_A)(x-\Omega_B)}
-\sum_{A=1}^{5}\frac{\p_6\p_A}{2a_0(x-\Omega_A)^2}
\left(1+\frac{3a_1(x-\Omega_A)}{a_0}\right).
\la{difur3}
\eea
It is useful to make the change of variables $u\to u(x)$. Then, performing
simple but tedious calculations which are outlined in Appendix B, one
obtains the following differential equation on $G(u)$ \bea
\partial_x \log G(u(x)) &=&
-\left(\frac {D}{16}+\frac{\p_6^2}{4}\right)\frac {d}{dx}\log u
+\frac {d_0}{x}
+\frac {d_1}{x-1}
+\frac {d_2}{x+\frac {n_0}{N-n_0}}\nonumber\\
&+&\frac {d_3}{x-\frac {N-n_0-n_\infty }{N-n_0}}
+\frac {d_4}{x-\frac {n_0}{n_0-n_\infty }}
-d_5\left(\frac {1}{x-\alpha_1}+\frac {1}{x-\alpha_2}\right) .
\la{difur4}
\eea
Here
$$
\alpha_i =\frac {n_0}{n_0-n_\infty }
+(-1)^i \sqrt {\frac {n_0n_\infty (N-n_\infty)}{(n_0-n_\infty )^2(N-n_0)}}
$$
are roots of the equation $x^2a_0=0$ and the coefficients $d_i$ are given by
the following formulas
\bea
\nonumber
d_0&=&\frac{D}{24}
\left(1-\frac{N-n_\infty}{n_0}-\frac{n_0}{N-n_\infty}\right)
+\frac{N-n_\infty}{2n_0}{\p_1^2}+\frac{n_0}{2(N-n_\infty)}{\p_4^2}
+\p_1\p_4+\p_6\p_1+\p_6\p_4
+\frac{\p_6^2}{2}, \\
\nonumber
d_1&=&
\frac{D}{24}\left(1+\frac{N-n_\infty}{n_\infty}+\frac{n_\infty}{N-n_\infty}\right)
-\frac{N-n_\infty}{2n_\infty}\p_3^2-\frac{n_\infty}{2(N-n_\infty)}\p_4^2
+\p_6\p_3+\p_6\p_4+\p_3\p_4+\frac{\p_6^2}{2},
\\
\nonumber
d_2&=&\frac{D}{24}\left(1+\frac{N-n_0}{n_0}+\frac{n_0}{N-n_0}\right)
-\frac{N-n_0}{2n_0}\p_1^2-\frac{n_0}{2(N-n_0)}\p_2^2
+\p_1\p_2+\p_6\p_1+\p_6\p_2+\frac{\p_6^2}{2}, \\
\nonumber
d_3&=&\frac{D}{24}
\left(1-\frac{n_\infty}{N-n_0}-\frac{N-n_0}{n_\infty}\right)
+\frac{n_\infty}{2(N-n_0)}\p_2^2+\frac{N-n_0}{2n_\infty}\p_3^2
+\p_6\p_2+\p_6\p_3+\p_2\p_3+\frac{\p_6^2}{2},
\\
\nonumber
d_4&=&\frac{D}{24}\left(1-\frac{n_0}{n_\infty}-\frac{n_\infty}{n_0}\right)
+\frac{n_\infty}{2n_0}\p_1^2+\frac{n_0}{2n_\infty}\p_3^2
+\p_6\p_1+\p_6\p_3+\p_1\p_3+\frac{\p_6^2}{2},
\\
\la{defd}
d_5&=&\frac{D}{24}-\frac{4\p_5\p_6-\p_5^2-\p_6^2}{4}.
\eea

Thus, with the account of the momentum conservation law
the solution of eq.(\ref{difur4}) is given by
\bea
G(u)&=&C(g_0,g_\infty)R^{D/2}
\frac{x^{d_0}(x-1)^{d_1}\left(x+\frac{n_0}{N-n_0}\right)^{d_2}
\left(x-\frac{N-n_0-n_\infty}{N-n_0}\right)^{d_3}
\left(x-\frac{n_0}{n_0-n_\infty}\right)^{d_4}
}
{u^{\frac{D}{16}+\frac{\p_6^2}{4}}
((x-\alpha_1)(x-\alpha_2))^{d_5}}.
\la{soldif}
\eea
Here $x=x(u)$ is the root of equation $u=u(x)$ that corresponds to given values
of the indices $I,J,K,L$, and $C(g_0,g_\infty )$ is a normalization constant
which does not depend on $u$.

Now we proceed with calculation of the correlation function $G^{ij}_b(u)$.
Setting in (\ref{soldif}) $D=8$, $\p_1=\k_1/2$,
$\p_2=\k_2/2$, $\p_3=\k_3/2$, $\p_4=\k_4/2$ and $\p_5=\p_6=0$
one obtains the bosonic correlation function $G_b(u)$ (\ref{cor})
up to the normalization constant $C(g_0,g_\infty )$.
Then, according to (\ref{Gbu1}),
in the limit $z\to 1$ and $w\to u$ we find
\bea
\nonumber
\lim_{z\to 1,w\to u}(z-1)^{1/2}(w-u)^{1/2} G_{IK}^{ij}(z,w)
=\frac{1}{4u^{1/2}(a_0(t_1)a_0)^{1/2}}\left(
-\d^{ij}\frac{1}{(t_1-x)^{2}}
-\sum_{AB}\frac{k_A^i k_B^j}
{4(t_1-\Omega_A)(x-\Omega_B)}\right),
\eea
where
\bea
a_0(t_1)=-\frac {1}{2}\left( \frac {n_0}{t_1^{2}}+
\frac {N-n_0}{(t_1-t_0)^{2}}-\frac
{N-n_\infty}{(t_1-t_\infty)^{2}}\right).
\nonumber
\eea
Taking into account that
$$
a_0(t_1)=-\frac{n_{\infty}^3(N-n_\infty)}{n_0(N-n_0)(n_0-n_\infty)^2}
\frac{\left(x+\frac{n_0}{N-n_0}\right)^2x^2}{(x-1)^2
\left(x-\frac{N-n_0-n_\infty}{N-n_0}\right)^2
\left(x-\frac{n_0}{n_0-n_\infty}\right)^2 }a_0,
$$
and performing simple calculations one obtains
\bea
\nonumber
&&G^{ij}_b(u)=
\frac{i}{8}
\left(\frac{n_{\infty} n_0 (N-n_{\infty})}{(N-n_0)}\right)^{1/2}
\frac{\l \t_i\t_j \r}{(n_{\infty}-n_0)u^{1/2}(x-\al_1)(x-\al_2)}G_b(u), 
\eea 
where we introduced a concise notation 
\bea 
&&
\la{tt}
\l 
\t_i\t_j\r = -\delta^{ij}\frac{4 x(x-1)\7 \8 \9 } 
{(n_0-n_{\infty})(x-\alpha_1)^2(x-\alpha_2)^2}
\\
\nonumber
&&
-\left(
\frac{\7}{n_0}k_1^i+\frac{\8}{n_\infty} k_3^i+\frac{1}{N-n_0}k_4^i
\right)
\left((x-1)k_1^j+x k_3^j
+\frac{n_0-n_{\infty}}{N-n_{\infty}}\9 k_4^j\right).
\eea
Thus, we determined the correlation function $G_b^{ij}(u)$ up to the
constant occuring in $G_b(u)$. However,
we are only interested in the overall normalization
constant for correlation function (\ref{guu}). To determine this constant
we are in need of the structure
constants in the OPE (\ref{OPE}). As will be shown later these
structure constants can be expressed through the corresponding
constants for the bosonic and fermionic twist fields. The latter
can be extracted from an auxiliary correlation function
\be
G_0(u)=
\langle\s_{ g_0^{-1}}[-\p_1,-\p_2](\infty )
\s_{IJ}[-\p](1)\s_{IJ}[\p](u)\s_{g_0}[\p_1,\p_2](0)\rangle ,
\la{gouu} \ee
where $I=1,\ldots ,n_0$, $J=n_0+1,\ldots ,N$.

\noindent Let us note that by using the action of $C_{g_0}$ one can fix
$I=n_0$, $J=N$. This correlation function corresponds to the case $n_\infty =n_0$ and the
rational function $u(x)$ is equal to
\be
u(x)=\left( 1+\frac {2n_0-N}{N-n_0}\frac {1}{x}\right)^{N-2n_0}
\left(\frac { 1+\frac {n_0}{N-n_0}\frac {1}{x}}{1-\frac {1}{x}}
\right)^{N}.
\la{uxo}
\ee
The root of eq.(\ref{uxo}) that corresponds to the correlation function (\ref{gouu})
behaves as
\be
\frac {1}{x}=\frac {1}{4n_0}(u-1) +o(u-1),\qquad \mbox{ when} \quad u\to 1.
\la{asym}
\ee
The following expression for the correlation function $G_0(u)$ can be derived from
eq.(\ref{soldif}) in the limit $n_\infty\to n_0$
\bea
G_0(u)&=&C(g_0)R^{D/2}
\frac{x^{d_0}(x-1)^{d_1}\left(x+\frac{n_0}{N-n_0}\right)^{d_2}
\left(x-\frac{N-2n_0}{N-n_0}\right)^{d_3}
}
{u^{\frac{D}{16}+\frac{\p^2}{4}}
\left(x-\frac {N-2n_0}{2(N-n_0)}\right)^{d_5}},
\la{soldif1}
\eea
where the coefficients $d_i$ are given by eq.(\ref{defd}) with the obvious
substitution $n_\infty\to n_0$, $\p_3=-\p_1$, $\p_4=-\p_2$, and
$\p_6=-\p_5=\p$.

Taking into account the OPE
\bea
\s_{IJ}[-\p](1)\s_{IJ}[\p](u)=
\frac{R^{-D/2}}{(1-u)^{\frac{D}{8}+\frac{\p^2}{2}}} +\cdots ,
\nonumber
\eea
and the normalization (\ref{norm}) of two-point correlation functions, one
gets
\be
G_0(u)\rightarrow
\frac{R^{-D/2}}{(1-u)^{\frac{D}{8}+\frac{\p^2}{2}}}.
\la{asgo1}
\ee

From the other side by using
eqs.(\ref{asym}) and (\ref{soldif1}), one derives in the limit $u\to 1$
\be
G_0(u)\rightarrow C(g_0)R^{D/2}
\left( \frac {1}{4n_0}(u-1)\right)^{-(d_0+d_1+d_2+d_3-d_5)}
=\frac {R^{D/2}}{(u-1)^{\frac{D}{8}+\frac{\p^2}{2}}}
C(g_0)(4n_0)^{{\frac{D}{8}+\frac{\p^2}{2}}}.
\la{asgo2}
\ee
Comparing eqs.(\ref{asgo1}) and (\ref{asgo2}), one finds the normalization
constant
\bea
C(g_0)=(-4n_0)^{-{\frac{D}{8}-\frac{\p^2}{2}}}.
\nonumber
\eea
Let us now consider the limit $u\to 0$. Taking into account the OPE
\bea
\s _{n_0N}[\p](u)\s _{g_0}[\p_1,\p_2](0)
&=&\frac{ C_{n_0N,g_0}^{g_{n_0N}g_0}(\p_1,\p_2,\p)}
{u^{\frac{D}{16}+\frac{\p^2}{4}+\D _{g_0}[\p_1,\p_2]-
\D_{g_{n_0N}g_0}[\p_1+\p_2+\p]}}
\s_{g_{n_0N}g_0}[\p_1+\p_2+\p](0)
\label{ope2}
\\
&+&
\frac{ C_{n_0N,g_0}^{g_0g_{n_0N}}(\p_1,\p_2,\p)}
{u^{\frac{D}{16}+\frac{\p^2}{4}+\D _{g_0}[\p_1,\p_2]-
\D_{g_{n_0N}g_0}[\p_1+\p_2+\p]}}
\s_{g_0g_{n_0N}}[\p_1+\p_2+\p](0)+\cdots ,
\nonumber \eea
one obtains
\bea
\nonumber
&&G_0(u)\rightarrow
\\
&&\frac{ C_{n_0N,g_0}^{g_{n_0N}g_0}(\p_1,\p_2,\p)}
{u^{\frac{D}{16}+\frac{\p^2}{4}+\D _{g_0}[\p_1,\p_2]-
\D_{g_{n_0N}g_0}[\p_1+\p_2+\p]}}
\langle\s_{ g_0^{-1}}[-\p_1,-\p_2](\infty )
\s_{n_0N}[-\p](1)\s_{g_{n_0N}g_0}[\p_1+\p_2+\p](0)\rangle
\nonumber\\ &&+ \frac{ C_{n_0N,g_0}^{g_0g_{n_0N}}(\p_1,\p_2,\p)}
{u^{\frac{D}{16}+\frac{\p^2}{4}+\D _{g_0}[\p_1,\p_2]-
\D_{g_{n_0N}g_0}[\p_1+\p_2+\p]}}
\langle\s_{ g_0^{-1}}[-\p_1,-\p_2](\infty )
\s_{n_0N}[-\p](1)\s_{g_0g_{n_0N}}[\p_1+\p_2+\p](0)\rangle .
\nonumber
\eea
By using the global $S_N$ invariance and
the obvious symmetry property of the structure constant
$$C_{n_0N,g_0}^{g_{n_0N}g_0}(-\p_1,-\p_2,-\p)=
C_{n_0N,g_0}^{g_{n_0N}g_0}(\p_1,\p_2,\p)$$
it is not difficult to show that the correlation functions
$\langle\s_{ g_0^{-1}}\s_{n_0N}\s_{g_{n_0N}g_0}\rangle$ and
$\langle\s_{ g_0^{-1}}\s_{n_0N}\s_{g_0g_{n_0N}}\rangle$ are equal to
$R^{D/2}C_{n_0N,g_0}^{g_0g_{n_0N}}$
and $R^{D/2}C_{n_0N,g_0}^{g_{n_0N}g_0}$ respectively:
\bea
\nonumber
&&\langle\s_{ g_0^{-1}}\s_{n_0N}\s_{g_{n_0N}g_0}\rangle =
\langle\s_{g_{n_0N}g_0}\s_{n_0N}\s_{ g_0^{-1}}\rangle
=\langle\s_{g_{n_0N}g_0^{-1}}\s_{n_0N}\s_{ g_0}\rangle
=R^{D/2}C_{n_0N,g_0}^{g_0g_{n_0N}}\\
&&\langle\s_{ g_0^{-1}}\s_{n_0N}\s_{g_0g_{n_0N}}\rangle =
\langle\s_{g_0g_{n_0N}}\s_{n_0N}\s_{ g_0^{-1}}\rangle
=\langle\s_{g_0^{-1}g_{n_0N}}\s_{n_0N}\s_{ g_0}\rangle
=R^{D/2}C_{n_0N,g_0}^{g_{n_0N}g_0}.
\label{2}
\eea
Moreover, the structure constants
$C_{n_0N,g_0}^{g_0g_{n_0N}}$
and $C_{n_0N,g_0}^{g_{n_0N}g_0}$ are complex-conjugated to each other.
Indeed, it follows from the normalization of the two-point
correlation functions of the twist fields $\s_g[\{\k_{\al}\}](z)$ that the
conjugate operator $(\s_g[\{\k_{\al}\}](z))^{\dagger}$ is given by
\be
(\s_g[\{\k_{\al}\}](z))^{\dagger}=
z^{-2\D_g[\{\k_{\al}\}]}\s_{g^{-1}}[\{-\k_{\al}\}]\left(\frac{1}{z}\right).
\la{dagger}
\ee
Then, from eqs.(\ref{2}) and (\ref{dagger}) one gets
$$
(R^{D/2}C_{n_0N,g_0}^{g_0g_{n_0N}})^*=
\langle\s_{ g_0^{-1}}\s_{n_0N}\s_{g_{n_0N}g_0}\rangle ^*=
\langle\s_{g_{n_0N}g_0}^{\dagger}\s_{n_0N}^{\dagger}\s_{g_0^{-1}}^{\dagger}
\rangle =
\langle\s_{g_0^{-1}g_{n_0N}}\s_{n_0N}\s_{ g_0}\rangle
=R^{D/2}C_{n_0N,g_0}^{g_{n_0N}g_0}.
$$
Thus, the correlation function $G_0(u)$ in the limit $u\to 0$ is expressed through
the structure constant
$$
C(n_0,\p_1;N-n_0,\p_2;\p)\equiv
C_{n_0N,g_0}^{g_{n_0N}g_0}(\p_1,\p_2,\p)
$$ as follows
\be
G_0(u)\rightarrow
\frac{2R^{D/2}|C(n_0,\p_1;N-n_0,\p_2;\p)|^2}
{u^{\frac{D}{16}+\frac{\p^2}{4}+\D _{g_0}[\p_1,\p_2]-
\D_{g_{n_0N}g_0}[\p_1+\p_2+\p]}}.
\la{asgo4}
\ee
On the other hand, taking into account that in the limit $u\to 0$ the root
$x(u)$ behaves as
\bea
x+\frac{n_0}{N-n_0}\to
-Nn_0^{\frac{N-2n_0}{N}}(N-n_0)^{\frac{2n_0-2N}{N}}u^{\frac{1}{N}},
\nonumber
\eea
one gets from eq.(\ref{soldif1})
\be
G_0(u,\bar u)\rightarrow 2^{\frac{1}{2}\p^2-\frac{5}{24}D}
\frac{R^{D/2}}{u^{\frac{D}{16}+\frac{\p^2}{4}-\frac{1}{N}d_2}}
\frac{
(N-n_0)^{-\frac{D}{24}+2\frac{n_0-N}{N}d_2+2\p\p_2+\frac{1}{2}\p^2}
n_0^{-\frac{D}{24}-2\frac{n_0}{N}d_2+2\p\p_1+\frac{1}{2}\p^2} }
{N^{\frac{D}{24}-2d_2+2\p(\p_1+\p_2)+\frac{3}{2}\p^2}}.
\la{asgo5}
\ee
Comparing eqs.(\ref{asgo4}) and (\ref{asgo5}), one obtains the following
expression for the modulus of the structure constant
\be
|C(n_0,\p_1;N-n_0,\p_2;\p)|=2^{-\frac{5D+24}{48}+\frac{1}{4}\p^2}
\frac{
(N-n_0)^{-\frac{D}{48}+\frac{n_0-N}{N}d_2+\p\p_2+\frac{1}{4}\p^2}
n_0^{-\frac{D}{48}-\frac{n_0}{N}d_2+\p\p_1+\frac{1}{4}\p^2} }
{N^{\frac{D}{48}-d_2+\p(\p_1+\p_2)+\frac{3}{4}\p^2}},
\la{strcons}
\ee
where
\bea
\nonumber
d_2&\equiv &d_2(n_0,\p_1;N-n_0,\p_2;\p)\\
&=&\frac{D}{24}\left(1+\frac{N-n_0}{n_0}+\frac{n_0}{N-n_0}\right)
-\frac{N-n_0}{2n_0}\p_1^2-\frac{n_0}{2(N-n_0)}\p_2^2
+\p_1\p_2+\p\p_1+\p\p_2+\frac{\p^2}{2}.
\nonumber
\eea
Note that the phase of the structure constant remains to be undetermined.

It is now not difficult to express any three-point correlation function of the form
$\langle\s_{ g^{-1}g_{IJ}}\s_{IJ}\s_{g}\rangle$ through the structure constant
$C(n,\k ;m,\q ;\p)$. Recall that any twist field
$\s_g[\{\k_\alpha\} ]$ has the following decomposition into the product
of the twist fields
$\s_{(n)}[\k ]$
\bea
\s_g[\{\k_\alpha\} ]=\prod_{\al =1}^{N_{str}} \, \s_{(n_\al )}[\k_\al ],
\la{1aa}
\eea
where the element $g$ has the decomposition $(n_1)(n_2)\cdots (n_{N_{str}})$.

\noindent Then, due to eq.(\ref{2}), the structure constant
$C(n,\k ;m,\q ;\p)$ with arbitrary $n$ and $m$ is equal to
\be C(n,\k ;m,\q;\p )=
R^{-D/2}\langle\s_{(-n-m)}[-\k -\q -\p](\infty )
\s_{IJ}[\p](1)\s_{(n)}[\k ]\s_{(m)}[\q ](0)\rangle , \la{2aa}
\ee where $I\in (n)$ and $J\in (m)$.

\noindent
By using eqs.(\ref{1aa}) and (\ref{2aa}), one can easily get the following
expression for the three-point correlation function
\bea
&&\langle\s_{ g^{-1}g_{IJ}}[\{\q_\alpha\} ](\infty )
\s_{IJ}[\p](1)\s_{g}[\{\k_\alpha\} ](0)\rangle =
\langle\s_{g}[\{\k_\alpha\} ](\infty )
\s_{IJ}[\p](1)\s_{ g^{-1}g_{IJ}}[\{\q_\alpha\} ](0)\rangle \nonumber\\
&&=\langle\s_{(-n_1-n_2)}[\q ]\prod_{\al =3}^{N_{str}}\s_{(-n_\al )}[\q_\al
] (\infty )\s_{IJ}[\p](1) \s_{(n_1)}[\k_1 ]\s_{(n_2)}[\k_2 ]
\prod_{\al =3}^{N_{str}}\s_{(n_\al )}[\k_\al ](0)\rangle\nonumber\\
&&=\prod_{\al =3}^{N_{str}}R^{-D/2}\d_R^D(\q_\al +\k_\al )
\langle\s_{(-n_1-n_2)}[\q ](\infty )\s_{IJ}[\p](1)
\s_{(n_1)}[\k_1 ]\s_{(n_2)}[\k_2 ](0)\rangle\nonumber\\
&&=C(n_1,\k_1;n_2,\k_2;\p)R^{-D/2}\d_R^D(\q +\k_1+\k_2+\p)
\prod_{\al =3}^{N_{str}}R^{-D/2}\d_R^D(\q_\al +\k_\al ),
\nonumber
\eea
where $I\in (n_1)$ and $J\in (n_2)$.

\noindent
It is now clear that the structure constant $C_{IJ,g}^{g_{IJ}g}$ in the OPE
of $\s_{IJ}$ and $\s_g$ is just equal to $C(n_1,\k_1;n_2,\k_2;\p)$,
and that the structure constant $C_{IJ,g^{-1}g_{IJ}}^{g^{-1}}$ (which
is complex conjugated to $C_{IJ,g^{-1}g_{IJ}}^{g_{IJ}g^{-1}g_{IJ}}$
due to eq.(\ref{dagger})) in the OPE
\bea
&&\s _{IJ}[\p](u)\s
_{g^{-1}g_{IJ}}[\{\q_\al \} ](0)
=\sum_{\q_1,\q_2}\frac{\d_{\q_1+\q_2-\q -\p ,0}}
{u^{\frac{D}{16}+\frac{\p^2}{4}+\D _{g^{-1}g_{IJ}}[\{\q_\al \}
]-\D _{g}[\{\q_\al \} ]}}
\label{ope6}\\
&&\times\left(
C_{IJ,g^{-1}g_{IJ}}^{g^{-1}}(\q_1,\q_2,\p ) \s _{g^{-1}}[\{\q_\al \}
](0)+\stackrel{*}{C}_{IJ,g^{-1}g_{IJ}}^{g^{-1}}(\q_1,\q_2,\p ) \s
_{g_{IJ}g^{-1}g_{IJ}}[\{\q_\al \} ](0)\right) +\cdots  \nonumber
\eea
is equal to
\bea
C_{IJ,g^{-1}g_{IJ}}^{g^{-1}}(\q_1,\q_2,\p
)=R^{-D/2}C(n_1,\q_1;n_2,\q_2;\p ).  
\nonumber
\eea
In particular, the structure constants
$C_{n_\infty N,g_0}^{g_{n_\infty N}g_0}$ and
$C_{n_0+n_\infty ,N;g_0}^{g_{n_0+n_\infty ,N}g_0}$, which
will be used to find the overall normalization constant for correlation function
(\ref{guu})
are given by
\bea
\la{5a}
&&C_{n_\infty N,g_0}^{g_{n_\infty N}g_0}(\k_1,\k_2,\p)
= R^{-D/2}C(n_\infty -n_0,\k_1;N-n_\infty ,\k_2;\p),\\
&&C_{n_0+n_\infty ,N;g_0}^{g_{n_0+n_\infty ,N}g_0}(\k_1,\k_2,\p)=
R^{-D/2}C(N-n_\infty -n_0,\k_1;n_\infty ,\k_2;\p).\nonumber
\eea

\section{Fermionic correlation functions}
\setcounter{equation}{0}
To find the contribution of the left-moving fermions to
the graviton scattering amplitude one has to compute
the following correlation function of four fermion twist fields:
\bea
\l \S_{g_\infty}^{i_3i_4}(\infty)\S_{IJ}^i(1)\S_{KL}
^j(u)\S_{g_0}^{i_1i_2}(0)\r .
\la{fc}
\eea
Computation of (\ref{fc}) will be again based on the stress-energy
tensor method and the conformal map (\ref{map}). Instead of
$N$ fermions $\theta_I(z)$ on the $z$-sphere, obeying twisted
boundary conditions around the points $0,1,\infty$ and $u$,
on the $t$-sphere one has one fermion $\theta(t)$ with the Ramond
boundary condition around the points $\Omega_A$.

It is well known that the Ramond fermions
are created from the NS sector by the standard spin fields 
(see, e.g., \cite{FMS}).  The simplest way to deal with correlation 
functions of the spin fields is to bosonize the fermions. To this end 
it is convinient to use the $SU(4)\times U(1)$ formalism \cite{GSW}.  
Recall that with respect to the $SU(4)\times U(1)$ subgroup
representations ${\bf 8_v}$, ${\bf 8_s}$ and ${\bf 8_c}$
are decomposed as
\bea
\nonumber
{\bf 8_s}\to {\bf 4_{1/2}}+{\bf \bar{4}_{-1/2}}, \qquad
{\bf 8_c}\to {\bf 4_{-1/2}}+{\bf \bar{4}_{1/2}}, \qquad
{\bf 8_v}\to {\bf 6_0}+{\bf 1_1}+{\bf 1_{-1}}.
\eea
The corresponding basis for the fermions $\theta^a$ and their
spin fields $\S^{\dot{a}}$ and $\S^{i}$
consistent with this decomposition is given by
\bea
\nonumber
&&\T^A=\frac{1}{\sqrt{2}}(\theta^A+i\theta^{A+4}), \qquad
\T^{\bar A}=\frac{1}{\sqrt{2}}(\theta^A-i\theta^{A+4}), \\
\nonumber
&&\CS^{\dot{A}}=\frac{1}{\sqrt{2}}(\S^{\dot{A}}+i\S^{\dot{A}+4}), \qquad
\CS^{\dot{\bar{A}}}=\frac{1}{\sqrt{2}}(\S^{\dot{A}}-i\S^{\dot{A}+4}), \\
\nonumber
&&S^{A}=\frac{1}{\sqrt{2}}(\S^{2A-1}+i\S^{2A}), \qquad
S^{\bar{A}}=\frac{1}{\sqrt{2}}(\S^{2A-1}-i\S^{2A}),
\eea
where $A=1,\ldots ,4$. Note that the spin fields $\S^4$ and
$\S^{\bar{4}}$ transform as ${\bf 1_1}$ and ${\bf 1_{-1}}$ respectively.

As usual, bosonization of the fermions and their twist fields
up to cocycles is realized in terms of four bosonic fields $\phi^A$ as
\bea
\nonumber
\T^A=\e^{iq_B^A\phi^B}, \qquad
\CS^{\dot{A}}=\e^{iq^{\A}_{B}\phi^B},\qquad
S^{A}=\e^{i\phi^A},
\eea
where the weights of the spinor representations ${\bf 8_s}$ and ${\bf 8_c}$
are given by
\bea
\nonumber
&&\q^1=\frac{1}{2}(-1,-1,1,1);~~~
\q^2=\frac{1}{2}(-1,1,-1,1);~~~
\q^3=\frac{1}{2}(1,-1,-1,1);~~~
\q^4=\frac{1}{2}(1,1,1,1); \\
\nonumber
&&
\q^{\dot 1}=\frac{1}{2}(-1,1,1,1);~~~
\q^{\dot 2}=\frac{1}{2}(-1,-1,-1,1);~~~
\q^{\dot 3}=\frac{1}{2}(1,1,-1,1);~~~
\q^{\dot 4}=\frac{1}{2}(1,-1,1,1).
\eea
The Cartan generators of
$SU(4)\times U(1)$ in the bosonized form look as $H^A=i\partial \phi^A$.

Clearly, bosonization of the fermions of the orbifold model
is achieved by introducing $4N$ bosonic fields and reads as
$$
\T_I^A(z)=\e^{iq_B^A\phi^B_I(z)}.
$$
Twist fields $\s_g$ creating twisted sectors for the fields
$\phi^A_I(z)$ are introduced in the same manner as in Sec.3
with the only exception that now they have the unit norm.
Since, $\s_{(n)}$ on the $z$-sphere corresponds to the identity operator
on the $t$-sphere, it is natural to assume that the
spin twist fields of the orbifold model can be realized as
\bea
\la{bosor}
&&\CS_{(n)}^{\dot{A}}(z)=
\e^{\frac{i}{n}\sum_{I\in (n)}q^{\A}_{B}\phi^B_I(z)}\s_{(n)}(z)=
\s_{(n)}[\q^{\A}](z),
\\
\nonumber
&&S_{(n)}^{A}(z)=\e^{\frac{i}{n}\sum_{I\in (n)}\phi^A_I(z)}\s_{(n)}(z)=
\s_{(n)}[\eb^{A}](z),
\eea
where $\eb^{A}$ is a weight vector of ${\bf 8_v}$ with
components $\d^{A}_{B}$.

Indeed, according to (\ref{confdim2}), the conformal
dimension of the field $\s_{(n)}[\eb^{A}]$ is
\bea
\nonumber
\D_n[\eb^{A}]=\frac{1}{6}\left(n-\frac{1}{n}\right)+\frac{(\eb^A)^2}{2n}
\eea
and analogously for $\s_{(n)}[\q^{\A}]$. Since
$(\eb^A)^2=(\q^{\A})^2=1$ these fields have the correct conformal
dimension (\ref{fd}) of a spin twist field.

By using the bosonization rule one can now establish that
the OPE of fermions with the twist fields (\ref{bosor})
coincides with (\ref{fOPE}) up to a sign. The correct sign
dependence should be restored by taking into account
cocycles of fermions and twist fields. Fortunately,
our method of calculation does not involve the knowledge of
these cocycles.

We proceed with considering correlation function (\ref{fc}).
In the $SU(4)\times U(1)$ basis correlation function  (\ref{fc})
reduces to correlation functions of the form (\ref{gencor}):
\bea
G_f(u)=
\langle \s_{g_\infty}[\p_3,\p_4](\infty )
\s_{IJ}[\p_5](1)\s_{KL}[\p_6](u) \s_{g_0}[\p_1,\p_2](0)\rangle ,
\la{fercor}
\eea
where a momentum $\p_i$ is now some weight vector $\pm\eb ^A$.
The computation of (\ref{fercor}) for general values of $\p$
was performed in the previous Section and the answer is given by
eq.(\ref{soldif}).  Recall that the structure constant
(\ref{strcons}) occuring in the OPE of twist fields was found up to a
phase that may depend on $\eb ^A$.

Some comments are in order.
It follows from (\ref{ope2}) that the basic
OPE's of the spin twist fields $S_{IJ}$ and  $S_{g_0}$
in the bosonized form looks as
\bea
S_{n_0N}^A(u)S_{g_0}^{BC}(0)
&=&\frac{ C_{n_0N,g_0}^{g_{n_0N}g_0}(\eb^A,\eb^B,\eb^C)}
{u^{\frac{1}{2}+\D _{g_0}[\eb^B,\eb^C]-
\D_{g_{n_0N}g_0}[\eb^A+\eb^B+\eb^C]}}
\s_{g_{n_0N}g_0}[\eb^A+\eb^B+\eb^C](0)
\label{opef}
\\
&+&
\frac{ C_{n_0N,g_0}^{g_0g_{n_0N}}(\eb^A,\eb^B,\eb^C)}
{u^{\frac{1}{2}+\D _{g_0}[\eb^B,\eb^C]-
\D_{g_{n_0N}g_0}[\eb^A+\eb^B+\eb^C]}}
\s_{g_0g_{n_0N}}[\eb^A+\eb^B+\eb^C](0)+\cdots ,
\nonumber
\eea
and
\bea
S_{n_0N}^A(u)S_{g_0}^{\bar{B}C}(0)
&=&\frac{ C_{n_0N,g_0}^{g_{n_0N}g_0}(\eb^A,-\eb^B,\eb^C)}
{u^{\frac{1}{2}+\D _{g_0}[\eb^B,\eb^C]-
\D_{g_{n_0N}g_0}[\eb^A-\eb^B+\eb^C]}}
\s_{g_{n_0N}g_0}[\eb^A-\eb^B+\eb^C](0)
\label{opef1}
\\
&+&
\frac{ C_{n_0N,g_0}^{g_0g_{n_0N}}(\eb^A,-\eb^B,\eb^C)}
{u^{\frac{1}{2}+\D _{g_0}[\eb^B,\eb^C]-
\D_{g_{n_0N}g_0}[\eb^A-\eb^B+\eb^C]}}
\s_{g_0g_{n_0N}}[\eb^A-\eb^B+\eb^C](0)+\cdots ,
\nonumber
\eea
Here the conformal dimension of $\s_{g_0g_{n_0N}}[\eb^A\pm\eb^B+\eb^C]$
is given by
$$
\D_{g_{n_0N}g_0}[\eb^A\pm\eb^B+\eb^C]=
\frac{N}{6}+\frac{1}{3N}+\frac{(\eb^A\pm\eb^B+\eb^C)^2-1}{2N}.
$$
Obviously, the norm of the vectors $\eb^+=\eb^A+\eb^B+\eb^C$
and $\eb^-=\eb^A-\eb^B+\eb^C$ can be equal to $3,5,9$ and to $1,3,5$
respectively.  When $(\eb^-)^2=1$ the field $\s[\eb^-]$ naturally
corresponds to the spin twist field $S^-$.
The appearence in (\ref{opef}), (\ref{opef1}) the
bosonic twist fields carrying integral momenta with non-identity norms
implies the existence of new spin twist fields whose bosonic
realizations (up to cocycles) are given by
$\s_{g_{n_0N}g_0}[\eb^{\pm}]$.
The origin of these fields becomes clear if one considers
excited states of a given twisted sector.

To discuss the excited states
it is convenient to use another basis for fermion fields
$\psi_I$:
\bea
\nonumber
\psi^a_I(z)=\frac{1}{\sqrt{n}}\sum_{J=1}^{n}e^{-\frac{2\pi i}{n}IJ}\theta^a_J(z)=
\sum_{m\in {\bf Z}}\theta^a_{nm-I}z^{\frac{I}{n}-m-\frac{1}{2}}
\eea
with the twisting property $\psi_I^a(e^{2\pi i}z)=-e^{\frac{2\pi i}{n}I}\psi_I^a(z)$.

By applying the operators $\psi_I^a$
to the vacuum state $|\S^{\dot{\mu}}\r$ of a twisted sector $(n)$
one obtains the excited states
$(\psi_{I}^{a}\S^{\dot{\mu}})(0)=\theta_{-I}^a|\S^{\dot{\mu}}\r$:
$$
\psi_I^a(z)|\S^{\dot{\mu}}\r =\frac{1}{z^{\frac{1}{2}-\frac{I}{n}}}
\theta_{-I}^a|\S^{\dot{\mu}}\r + \mbox{reg}.
$$
The conformal dimension of the corresponding primary fields
$ \psi_{I}^{a}\S^{\dot{\mu}}$ is
$\D[\psi_{I}\S]=\frac{n}{6}+\frac{1}{3n}+\frac{I}{n}$.
In the same manner one can introduce primary fields
$ \psi_{I_1}^{a_1}\cdots \psi_{I_k}^{a_k}\S^{\dot{\mu}}$
corresponding to more general excited states of a twisted sector.
If $I_s+I_p\neq n$ for any $s$ and $p$, then their conformal dimensions are
given by
\be
\D[\psi_{I_1}\cdots \psi_{I_k}\S]=
\frac{n}{6}+\frac{1}{3n}+
\frac{I_1+\cdots+I_k}{n}.
\la{ecd}
\ee

Let us now consider the OPE of the $Z_2$-twist field $\S^i$
with the twist field $\S^{jk}$ corresponding to an element $g_0$.
The product $\S^i\S^{jk}$ transforms as the tensor product of
three ${\bf 8}_v$ representations of $SO(8)$ and, therefore, it
can be decomposed in a set of irreducible representations,
each of them realized by excited fields of a twisted sector.
Thus, schematically, the first few terms of the OPE
required by the $SO(8)$ symmetry read as
\bea
\la{invOPE}
\S^i(u)\cdot \S^{jk}(0)
&=&\d^{ij}\S^k+\d^{ij}\g^{k}_{a\a}\psi^a_I\S^{\a}+
\g^{ij}_{ab}\g^{k}_{b\a}\psi^a_I\S^{\a}
+\d^{ij}\g^{km}_{ab}\psi^a_I\psi^b_J\S^{m}+
\g^{ij}_{ab}\psi^a_I\psi^b_J\S^{k} \\
\nonumber
&+&
\d^{ij}\g^{km}_{ab}\g^{m}_{c\c}\psi^a_I\psi^b_J\psi^c_M\S^{\c}+
\g^{ij}_{ab}\g^{k}_{c\c}\psi^a_I\psi^b_J\psi^c_M\S^{\c}
+
\d^{ij}\g^{km}_{ab}\g^{mp}_{cd}\psi^a_I\psi^b_J\psi^c_M\psi^d_N\S^{p}\\
\nonumber
&+&
\g^{ij}_{ab}\g^{kp}_{cd}\psi^a_I\psi^b_J\psi^c_M\psi^d_N\S^{p}
+\g^{ip}_{ab}\g^{jp}_{cd}\psi^a_I\psi^b_J\psi^c_M\psi^d_N\S^{k}
+\mbox{cycl. perm.}~(i,j,k)+\ldots.
\eea
For the sake of simplicity  we omitted here the $SO(8)$ index
independent structure constants standing by each summond on the
r.h.s., as well as the $u$-dependence. Note that in (\ref{invOPE}) we
can take into account only nonzero indices $I,J,M,N$ since the action
of $\psi_0$ on a vacuum state does not produce a new excited field
and, therefore, does not disturb the form of the OPE.

Comparing $\D_{g_{n_0N}g_0}[\eb^{\pm}]$
with (\ref{ecd}) one can relate the norm of the vectors
$\eb^{\pm}$ with the numbers $I_s$ of excited states in
the twisted sector $(N)$:
\bea
\la{numb}
I_1+\cdots+I_k=\frac{(\eb^{\pm})^2-1}{2}=\{0,1,2,4\},
\eea
where on the r.h.s. all possible values of the sum $I_1+\cdots+I_k$
are indicated.

Under the $SU(4)\times U(1)$ group the $\g$-matrices $\g^{ij}$ are
decomposed into the matrices $\g^{AB}$, $\g^{A\bar{B}}$ and
$\g^{\bar{A}\bar{B}}$.
A simple analysis shows that in the $SU(4)\times U(1)$ basis
the invariant OPE (\ref{invOPE}) acquires a form
\bea
\la{mst1}
S_{n_0N}^A(u)S_{g_0}^{BC}(0)&=&\frac{c_3}{u^{\D_2}}
\g^{AB}_{ab}\g^C_{b\b}\psi_1^{a}\S^{\b}\\
\nonumber
&+&\frac{1}{u^{\D_3}}
\left(
c_5^{(1)}\g^{AB}_{ab}\psi_1^{a}\psi_1^{b}S^{C} +
c_5^{(2)}\g^{AC}_{ab}\psi_1^{a}\psi_1^{b}S^{B}+
c_5^{(3)}\g^{CB}_{ab}\psi_1^{a}\psi_1^{b}S^{A}
\right)\\
\nonumber
&+&\frac{c_9}{u^{\D_4}}
\g^{As}_{[ab}\g^{As}_{cd]}\psi_1^{a}\psi_1^{b}\psi_1^{c}\psi_1^{d}S^{A}
\\
\la{mst2}
S_{n_0N}^A(u)S_{g_0}^{\bar{B}C}(0)&=&\frac{1}{u^{\D_1}}
\left(c_1^{(1)}\d^{AB}S^C +c_1^{(3)}\d^{BC}S^A \right)
+\frac{c_3}{u^{\D_2}}\g^{A\bar{B}}_{ab}\g^C_{b\b}\psi_1^{a}\S^{\b}
+\frac{c_5}{u^{\D_3}}\g^{A\bar{B}}_{ab}\psi_1^{a}\psi_1^{b}S^{A},
\eea
where the coefficients $\D_1,\ldots,\D_4$ are defined by the conformal
symmetry and at the moment are unessential.

On the r.h.s. of eqs.(\ref{mst1}) and (\ref{mst2}) we have indicated
only the leading singular terms of the OPE's.
Now it is readily seen that they are in one to one correspondence
with bosonic fields $\s [\eb^{+}]$ arising on the r.h.s. of (\ref{opef}).
Consider, for instance, eq.(\ref{mst1}).
Due to the facts that $\g^{AB}=-\g^{BA}$ and $\g^A(\g^{A})^T=0$ for any
$A$, the first term in (\ref{mst1}) is nonzero only if $A\neq B\neq C$.
Then, according to (\ref{numb}), the  operators
$\g^{AB}_{ab}\g^C_{b\b}\psi_1^{a}\S^{\b}$
can be identified with the fields $\s[\eb^+]$ with $(\eb^+)^2=3$.
Next, when, e.g., $A=B\neq C$, the first nonzero term stands by
the singularity $\frac{1}{u^{\D_3}}$ and, therefore admits the
identification with $\s[\eb^+]$ with $(\eb^+)^2=5$.
The last term in (\ref{mst1}) becomes involved in the case $A=B=C$
and the corresponding operators can be regarded as
$\s[\eb^+]$ with $(\eb^+)^2=9$. The same situation
takes place for (\ref{mst2}), where the three singular terms
correspond to $\s[\eb^-]$ with $(\eb^-)^2=1,3,5$.
Thus, the structure constants in (\ref{opef}) and (\ref{opef1})
correspond to the ones in (\ref{mst1}) and (\ref{mst2}).

\section{Normalization of four-point correlation functions}
\setcounter{equation}{0}
In this section we combine the results obtained in the two
previous Sections and calculate the normalization constant
occuring in the product of bosonic and fermionic correlation functions:
\bea
\la{holcor}
G(u)=G_b(u)G_f(u)&=&
\langle \s_{g_\infty}[\k_3/2,\k_4/2](\infty )
\s_{IJ}(1)\s_{KL}(u) \s_{g_0}[\k_1/2,\k_2/2](0)\rangle
\\
\nonumber
&\times &\langle \s_{g_\infty}[\p_3,\p_4](\infty )
\s[\p_5]_{IJ}(1)\s[\p_6]_{KL}(u) \s_{g_0}[\p_1,\p_2](0)\rangle ,
\eea
where each $\p_i$ coincides with some $\pm\eb^A$.
According to (\ref{soldif}), one gets for $G(u)$:
\bea
G(u)&=&C(g_0,g_\infty)R^{4}
\frac{x^{d_0}(x-1)^{d_1}\left(x+\frac{n_0}{N-n_0}\right)^{d_2}
\left(x-\frac{N-n_0-n_\infty}{N-n_0}\right)^{d_3}
\left(x-\frac{n_0}{n_0-n_\infty}\right)^{d_4}
}
{u((x-\alpha_1)(x-\alpha_2))^{d_5}}.
\la{soldifg}
\eea
Here the coefficients $d_i$ are given by
\bea
\nonumber
d_0&=&
1+\frac{1}{4}k_1k_4+
\p_1\p_4+\p_6\p_1+\p_6\p_4,
\qquad
d_1=
1+\frac{1}{4}k_3k_4+\p_6\p_3+\p_6\p_4+\p_3\p_4,
\\
\nonumber
d_2&=&
1+\frac{1}{4}k_1k_2+\p_1\p_2+\p_6\p_1+\p_6\p_2,
\qquad
d_3=
1+\frac{1}{4}k_2k_3+\p_6\p_2+\p_6\p_3+\p_2\p_3,
\\
\nonumber
d_4&=&
1+\frac{1}{4}k_1k_3+\p_6\p_1+\p_6\p_3+\p_1\p_3,
\qquad
d_5=1-\p_5\p_6,
\nonumber
\eea
where $k_ik_j=\k_i\k_j-\frac{1}{2}k_i^+k_j^--\frac{1}{2}k_i^-k_j^+$.

The normalization constant $C(g_0,g_\infty )$ can be determined
by factorizing $G(u)$ in the limit $u\to 0$ on three-point
functions.
According to eq.(\ref{ux}), $u\to 0$ in the following
three cases
\bea
\nonumber
I)~~x\to -\frac{n_0}{N-n_0};~~~~
II)~~x\to \infty;~~~~
III)~~x\to \frac{N-n_0-n_{\infty}}{N-n_0}
\eea
and, conversely, any root $x_M=x_M(u)$ of eq.(\ref{ux}) tends to one of
these values when $u\to 0$. Evidently, these three possible
asymptotics correspond to three different choices of the indices $K$
and $L$ in eq.(\ref{s2}).

Let us begin with the case $K=n_0$, $L=N$. By using the OPE
(\ref{ope2}) and the normalization (\ref{norm}) of two-point correlation functions, one
gets in the limit $u\to 0$
\be
G(u)\rightarrow R^{4}
\frac{C(n_0,\k_1,\p_1;N-n_0,\k_2,\p_2;\p_6)
\stackrel{*}{C}(n_\infty ,\k_3,\p_3;N-n_\infty ,\k_4,\p_4;\p_5)}
{u^{1-\frac{d_2}{N}}},
\la{7a}
\ee
where we have taken into account that
$$
\D^b_{g_0}[\k_1,\k_2]+\D^f_{g_0}[\p_1,\p_2]
-\D^b_{g_{n_0N}g_0}[\k_1+\k_2]-\D^f_{g_{n_0N}g_0}[\p_1+\p_2+\p_6]
=-\frac{1}{N}d_2 ,
$$
Here $C(n_0,\k_1,\p_1;N-n_0,\k_2,\p_2;\p)$
denotes the product of the bosonic and fermionic structure
constants
$$
C(n_0,\k_1,\p_1;N-n_0,\k_2,\p_2;\p)=
C(n_0,\k_1/2;N-n_0,\k_2/2;0)C(n_0,\p_1;N-n_0,\p_2;\p).
$$
Due to (\ref{strcons}) the modulus of this structure constant
is equal to
\bea
\nonumber
|C(n_0,\k_1,\p_1;N-n_0,\k_2,\p_2;\p)|=
2^{-\frac{3}{2}}
\frac{
(N-n_0)^{\frac{n_0-N}{N}d_2+\p\p_2}
n_0^{-\frac{n_0}{N}d_2+\p\p_1}
}
{N^{1-d_2+\p(\p_1+\p_2)}},
\eea
where
\bea
\nonumber
d_2&\equiv &d_2(n_0,\k_1,\p_1;N-n_0,\k_2,\p_2;\p)
=1+\frac{1}{4}\left({\bf k}_1{\bf k}_2-\frac{N-n_0}{2n_0}{\bf k}_1^2
-\frac{n_0}{2(N-n_0)}{\bf k}_2^2\right)\\
&+&\frac{N-n_0}{2n_0}(1-\p_1^2)+\frac{n_0}{2(N-n_0)}(1-\p_2^2)+
\p_1\p_2+\p\p_1+\p\p_2.
\nonumber
\eea

The root
$x(u)$ has the following behaviour in the vicinity of $u=0$ \be
|x+\frac{n_0}{N-n_0}|\to
Nn_0^{\frac{N-n_0}{N}}n_\infty^{-\frac{n_\infty}{N}}
(N-n_0)^{\frac{n_0-2N}{N}}(N-n_\infty )^{\frac{n_\infty }{N}}|u|^{\frac{1}{N}}.
\la{8a}
\ee
By using eqs.(\ref{soldifg}) and (\ref{8a}), one can easily find
\bea
G(u)\to
\e^{i\varphi}\frac{C(g_0,g_{\infty})}{u^{1-\frac{{\bf d}_2}{N} }   }
\frac{n_0^{d_0+d_4+\frac{N-n_0}{N}d_2-1+\p_5\p_6}
N^{d_1+d_2-1+\p_5\p_6}
(N-n_\infty)^{d_2+d_3+d_4-\frac{N-n_\infty}{N}d_2
-1+\p_5\p_6}}
{ n_\infty^{\frac{n_\infty}{N}d_2}
(N-n_0)^{d_0+d_1+d_3+d_4+\frac{2N-n_0}{N}d_2
-2+2\p_5\p_6 }
(n_\infty-n_0)^{d_4-1+\p_5\p_6}},
\la{9a}
\eea
where the phase multiplier $\e^{i\varphi}$ depending, in particular,
on a phase behaviour of the root $x(u)$ remains to be undetermined.

\noindent Comparing eqs.(\ref{7a}) and (\ref{9a}), one obtains the
modulus of the normalization constant:
\bea
\la{12a}
|C(g_0,g_{\infty})|=2^{-3}
\frac{n_0^{\p_1\p_5}n_\infty^{\p_3\p_5}
(N-n_\infty)^{\p_4\p_6}
(n_\infty-n_0)^{\frac{1}{4}k_1k_3+\p_1\p_6+\p_3\p_6+\p_1\p_3+\p_5\p_6} }
{(N-n_0)^{1+\frac{1}{4}k_1k_3+\p_1\p_5+\p_3\p_5+\p_1\p_3-\p_2\p_6}}.
\eea
Thus, we have found the normalization constant up to a phase
for $N$ correlation functions which are
presented in the first and second terms of eq.(\ref{s2}).

Let us now determine the normalization constant for $n_\infty -n_0$ correlation functions
of the form  \\
$\langle V_{g_\infty (J)}
V_{n_0J}V_{n_\infty N}V_{g_0}\rangle $. By using the OPE (\ref{ope6}) and
eq.(\ref{5a}), one finds in the limit $u\to 0$
\bea
\nonumber
G(u)&\rightarrow &\frac{R^4
\stackrel{*}{C}(n_\infty-n_0,\k_1+\k_3,\p_1+\p_5+\p_3;N-n_\infty,\k_4,\p_4;\p_6)
}
{u^{1-\frac{1}{n_\infty-n_0}(1+\frac{1}{4}\k_1\k_3+\p_1\p_3+\p_3\p_5+\p_1\p_5)}}
\times \\
\nonumber
& \times &
C(n_\infty-n_0,-\k_1-\k_3,-\p_1-\p_5-\p_3;n_0,\k_1,\p_1;\p_5).
\eea
Taking into account the behaviour of the root $x(u)$ in the vicinity of $u=0$
\bea
|x|\to \left( (n_\infty -n_0)^{n_0-n_\infty}\frac{n_{\infty}^{n_{\infty}}}
{n_0^{n_0}}\left(\frac{N-n_0}{N-n_{\infty}}\right)^{N-n_{\infty}}\right)
^{\frac{1}{n_\infty -n_0}}|u|^{\frac{1}{n_0-n_\infty }},
\nonumber
\eea
one obtains from eq.(\ref{soldifg})
\bea
\nonumber
G(u)&\rightarrow &\frac{R^4C(g_0,g_\infty )}
{u^{1-\frac{1}{n_\infty-n_0}(1+\frac{1}{4}\k_1\k_3+\p_1\p_3+\p_3\p_5+\p_1\p_5)}}
\\
\nonumber
&\times
&\left( (n_\infty -n_0)^{n_0-n_\infty}\frac{n_{\infty}^{n_{\infty}}}
{n_0^{n_0}}\left(\frac{N-n_0}{N-n_{\infty}}\right)^{N-n_{\infty}}\right)
^{-\frac{1+\frac{1}{4}\k_1\k_3+\p_1\p_3+\p_3\p_5+\p_1\p_5}{n_\infty-n_0}},
\eea
where we have used the relation
$$
d_0+d_1+d_2+d_3+d_4+2d_5=-\left(1+\frac{1}{4}k_1k_3
+\p_1\p_3+\p_1\p_5+\p_3\p_5\right).
$$
Now to find the normalization
constant one should take into account the following identities:
\bea
\nonumber
&&d_2(n_\infty-n_0,\k_1+\k_3,\p_1+\p_3+\p_5;N-n_\infty,\k_4,\p_4;\p_6)
=-\frac{N-n_0}{n_\infty-n_0}(1+\frac{1}{4}k_1k_3+\p_1\p_3+\p_1\p_5+\p_2\p_5)\\
\nonumber
&&d_2(n_\infty-n_0,-\k_1-\k_3,-\p_1-\p_3-\p_5;n_0,\k_1,\p_1;\p_5)
=-\frac{n_\infty}{n_\infty-n_0}(1+\frac{1}{4}k_1k_3+\p_1\p_3+\p_1\p_5+\p_2\p_5)
\eea
By using these formulae one can establish that the modulus
of the normalization constant is again given by (\ref{12a}).

The normalization constant for the remaining $N-n_0-n_\infty$ correlation functions of
the form
$$\langle V_{g_\infty (J)}V_{n_0J}V_{n_0+n_\infty ,N}V_{g_0}\rangle $$
can be found in the same manner and is again defined by
eq.(\ref{12a}).

Now we are ready to establish relative signs of the normalization
constants corresponding to different values of $\p$'s.
Note that the $SO(8)$ invariance of the model dictates the form
of the correlation function
\bea
\nonumber
G^{i_1i_2i_3i_4i_5i_6}_{IJKL}(u)=
\langle V_{g_\infty}^{i_3i_4}[\k_3,\k_4](\infty)
V^{i_5}_{IJ}(1) V^{i_6}_{KL}(u)
V^{i_1i_2}_{g_0}[\k_1,\k_2](0)\rangle ,
\eea
where we have used the notation
$V^{i_1i_2}_{g_0}[\k_1,\k_2]=\left(\s[\k_1/2,\k_2/2]\S^{i_1i_2}\right)_{g_0}$.

Namely, this correlation function is decomposed into $SO(8)$ invariant
rank six tensors:
\bea
\nonumber
G^{i_1i_2i_3i_4i_5i_6}(u)=
C_1(u)\d^{i_1i_2}\d^{i_3i_4}\d^{i_5i_6}+C_2(u)\d^{i_1i_2}\d^{i_3i_6}\d^{i_4i_5}
+\cdots .
\eea
Here the total number of terms is equal to 15.
In fact, each function $C_i(u)$ coincides up to a phase with the
correlation function (\ref{holcor}) under a particular choice of $\p$'s.
For example, by using the $SU(4)\times U(1)$ basis,
the functions $C_1(u)$ and $C_2(u)$ can be
schematically expressed as
\bea
&&C_1(u)=\l \s S^{B\bar{B}}\s S^{C}\s S^{\bar{C}}\s S^{A\bar{A}}\r
\sim
\l \s[\eb^B,-\eb^B]\s[\eb^C]\s[-\eb^C]\s[\eb^A,-\eb^A]\r
\\
\nonumber
&&C_2(u)=\l \s S^{B\bar{C}}\s S^{C}\s S^{\bar{B}}\s S^{A\bar{A}}\r
\sim
\l \s[\eb^B,-\eb^C]\s[\eb^C]\s[-\eb^B]\s[\eb^A,-\eb^A]\r ,
\eea
where one has to choose all the vectors $\eb^A$, $\eb^B$ and $\eb^C$
to be different. On the other hand, if $C=B$, then one can recognize
that
\bea
C_1(u)+C_2(u)=
\l \s S^{B\bar{B}}\s S^{B}\s S^{\bar{B}}\s S^{A\bar{A}}\r
\sim
\l \s[\eb^B,-\eb^B]\s[\eb^B]\s[-\eb^B]\s[\eb^A,-\eb^A]\r .
\nonumber
\eea
Since we know all three correlation functions up to phases we get a nontrivial
relation on $C_1(u)$ and $C_2(u)$ allowing one to determine their relative
sign. Namely, from (\ref{soldifg}) with the account of the found
normalization constants (\ref{12a}) one obtains
\bea
\nonumber
&&C_1(u)\sim \frac{\e^{i\varphi_1}}{(n_\infty-n_0)\6\7
(x-\al_1)(x-\al_2)}, \\
\nonumber
&&C_2(u)\sim -\frac{\e^{i\varphi_2}}{(n_\infty-n_0)\6\7\8\9 }, \\
\nonumber
&&C_1(u)+C_2(u)
\sim \frac{\e^{i\varphi_3}(N-n_\infty)n_\infty x}
{(N-n_0)(n_\infty-n_0)^2\6\7\8\9 (x-\al_1)(x-\al_2)},
\eea
where a common multiplier coming in all three functions is omitted.

Now it can be readily seen that the last equation is satisfied
only if $\varphi_1=\varphi_2=\varphi_3=\varphi$.
Proceeding in the same manner we fix the relative
signs of all 15 functions $C_i$. The final answer for the correlation function
$G^{i_1i_2i_3i_4ij}_{IJKL}(u)$ reads as
\bea
\la{funG}
&&G^{i_1i_2i_3i_4ij}_{IJKL}(u)=
-\e^{i\varphi}\frac{R^4}{8(N-n_0)}
\left(\frac{n_\infty-n_0}{N-n_0}\right)^{\frac{1}{4}k_1 k_3}\\
\nonumber
&&
\times \frac{\9 ^3}{u(x-\al_1)(x-\al_2)}
\left(\frac{x\8}{\9}\right)^{1+\frac{1}{4}k_1k_4}
\left(\frac{(x-1)\7}{\9}\right)^{1+\frac{1}{4}k_3k_4}
T_{IJKL}^{i_1i_2i_3i_4 ij}(u),
\eea
where 
\bea
\nonumber
&&T_{IJKL}^{i_1i_2i_3i_4 ij}(u)=\\
\nonumber
&&=
\frac{\d^{ij}}{(n_0-n_\infty)(x-\alpha_1)(x-\alpha_2)}
\left(
\frac{\d^{i_1i_2}\d^{i_3i_4}}{\6\7}
-\frac{\d^{i_1i_4}\d^{i_2i_3}}{\5\8}
-\frac{N-n_0}{(n_0-n_\infty)}\frac{\d^{i_1i_3}\d^{i_2i_4}}{\9}
\right)
\\
\nonumber
&&+\frac{\d^{i_3i_4}}{\6\7}
\left(\frac{\d^{i i_1 }\d^{j i_2}}{n_0\8}-
\frac{\d^{i i_2 }\d^{j i_1}}{(n_0-n_\infty)\5\9}
\right)\\
\nonumber
&&+\frac{\d^{i_1 i_2 }}{\6\7}
\left(
\frac{N-n_0}{n_\infty(N-n_\infty)}
\frac{\d^{i i_3 }\d^{j i_4}}{\5}
-\frac{\d^{i i_4 }\d^{j i_3}}{(n_0-n_\infty)\8\9}
\right)\\
\nonumber
&&+\frac{\d^{i_2 i_3}}{\5\8}
\left(
\frac{\d^{i i_4 }\d^{j i_1}}{(n_0-n_\infty)\7\9}-
\frac{N-n_0}{n_0(N-n_\infty)}
\frac{\d^{i i_1}\d^{j i_4}}{\6}
\right)\\
\nonumber
&&+\frac{\d^{i_2 i_4}}{\9}
\left(
\frac{N-n_0}{n_0(n_0-n_\infty)}\frac{\d^{i i_1}\d^{j i_3}}{\6\8}-
\frac{N-n_0}{n_\infty(n_0-n_\infty)}
\frac{\d^{i i_3 }\d^{j i_1}}{\5\7}
\right) \\
\nonumber
&&+\frac{\d^{i_1 i_4}}{\5\8}
\left(
\frac{\d^{i i_2}\d^{j i_3}}{(n_0-n_\infty)\6\9}-
\frac{\d^{i i_3}\d^{j i_2}}{n_\infty\7}
\right) \\
\nonumber
&&+\frac{\d^{i_1 i_3}}{\9} 
\left(
\frac{N-n_0}{(N-n_\infty)(n_0-n_\infty)}\frac{\d^{i i_2}
\d^{j i_4}}{\5\6}-
\frac{\d^{i i_4}\d^{j i_2}}{(n_0-n_\infty)\7\8}
\right).
\eea

Certainly, the common phase $\varphi$
remains to be undetermined and can depend on the indices $I,J,K,L$
and the momenta $k_i$. However, we will see at a moment that
this phase disappears if one takes into account the contribution of
the right-moving sector.
By using the world-sheet parity symmetry combined with space reflection
(\ref{trans}) we get the following relation between the correlation functions
of the holomorphic and antiholomorphic sectors:
\bea
\nonumber
&&\langle V_{g_\infty}^{i_3i_4}[\k_3,\k_4](\infty)
V^{i_5}_{IJ}(1) V^{i_6}_{KL}(u)
V^{i_1i_2}_{g_0}[\k_1,\k_2](0)\rangle ^*\\
&&=
\nonumber
\langle
\bar{V}^{\tilde{i}_3\tilde{i}_4}_{g_\infty^{-1}}
[\tilde{\k}_3,\tilde{\k}_4](\infty )
\bar{V}^{\tilde{i}_5}_{IJ}(1)
\bar{V}^{\tilde{i}_6}_{KL}(u)
\bar{V}^{\tilde{i}_1\tilde{i}_2}_{g_0^{-1}}
[\tilde{\k}_1,\tilde{\k}_2](0)\rangle ^*\\
\nonumber
&&=u^{-2}
\langle
\bar{V}^{\tilde{i}_1\tilde{i}_2}_{g_0}[-\tilde{\k}_1,-\tilde{\k}_2](\infty)
\bar{V}^{\tilde{i}_6}_{KL}(u^{-1})
\bar{V}^{\tilde{i}_5}_{IJ}(1)
\bar{V}^{\tilde{i}_3\tilde{i}_4}_{g_\infty}[-\tilde{\k}_3,-\tilde{\k}_4](0)
\rangle \\
\nonumber
&&=
\langle
\bar{V}^{\tilde{i}_3\tilde{i}_4}_{g_\infty}[-\tilde{\k}_3,-\tilde{\k}_4](\infty )
\bar{V}^{\tilde{i}_5}_{IJ}(1)
\bar{V}^{\tilde{i}_6}_{KL}(u)
\bar{V}^{\tilde{i}_1\tilde{i}_2}_{g_0}[-\tilde{\k}_1,-\tilde{\k}_2](0)
\rangle \\
\nonumber
&&=\langle \bar{V}_{g_\infty}^{i_3i_4}[\k_3,\k_4](\infty)
\bar{V}^{i_5}_{IJ}(1) \bar{V}^{i_6}_{KL}(u)
\bar{V}^{i_1i_2}_{g_0}[\k_1,\k_2](0)\rangle ,
\eea
and we recall that
$V^{i_1i_2}_{g_0}[\k_1,\k_2]=\left(\s[\k_1/2,\k_2/2]\S^{i_1i_2}\right)_{g_0}$
and
$\bar{V}^{i_1i_2}_{g_0}[\k_1,\k_2]=
\left(\bar{\s}[\k_1/2,\k_2/2]
\bar{\S}^{i_1i_2}\right)_{g_0}$.
Here the conjugation property (\ref{dagger})
of $V$ and the invariance of the correlation function (\ref{funG}) under the space
reflection were used. Thus, correlation functions of the anti-holomorphic
sector are complex-conjugated to correlation functions of the holomorphic one.
Therefore, after combining these two sectors the phase ambiguity
dissappears.

\section{Scattering amplitude}
\setcounter{equation}{0}
The results obtained in the previous section allows one
to determine a holomorphic contribution $G_{IJKL}^{i_1i_2i_3i_4}(u)$
to the correlation function (\ref{guu})
\bea
\la{GIJKL}
G_{IJKL}^{i_1i_2i_3i_4}(u)&=&G_b^{ij}(u)\l \S_{g_\infty}^{i_3i_4}(\infty)\S_{IJ}^i(1)\S_{KL}
^j(u)\S_{g_0}^{i_1i_2}(0)\r \\
&=&
\nonumber
\frac{i}{8}
\left(\frac{n_{\infty} n_0 (N-n_{\infty})}{(N-n_0)}\right)^{1/2}
\frac{x\l \t_i\t_j\r G^{i_1i_2i_3i_4ij}_{IJKL}(u)}
{(n_{\infty}-n_0)u^{1/2}(x-\al_1)(x-\al_2)}.
\eea

Up to now we considered the correlation functions
$G_{IJKL}^{i_1i_2i_3i_4}(u)$ with $|u|<1$. The correlation functions $G_{IJKL}^{i_1i_2i_3i_4}(u)$ with
$|u|>1$ can be calculated in the same way.
In particular, the normalization constant in this case is
derived by studying the limit $u\to\infty$ and
coincides with the previously found
constant (\ref{12a}). We find that $G_{IJKL}^{i_1i_2i_3i_4}(u)$ with $|u|<1$
is again given  by eq.(\ref{GIJKL}).
The time-ordering, therefore, can be omitted, and to
complete the computation of the S-matrix element we have to integrate the
product $G_{IJKL}^{i_1i_2i_3i_4}(u)\bar{G}_{IJKL}^{j_1j_2j_3j_4}(\bar{u})$ over the complex plane.
With the account of eq.(\ref{funG}) and the equality
\bea
\frac {1}{u}\frac {du}{dx}=
\frac {(n_0-n_\infty )(x-\al_1)^2(x-\al_2)^2}
{x(x-1)(x-\frac {N-n_0-n_\infty }{N-n_0})
(x-\frac {n_0}{n_0-n_\infty })(x+\frac {n_0}{N-n_0})}\nonumber
\eea
we find that the corresponding integral is given by
\bea
\la{int}
&&\int d^2u |u| G_{IJKL}^{i_1i_2i_3i_4}(u)\bar{G}_{IJKL}^{j_1j_2j_3j_4}(\bar{u})
=\kappa\frac{R^8}{2^{12}}\frac{n_0n_\infty(N-n_\infty)}{(N-n_0)^3}
\left(\frac{n_\infty-n_0}{N-n_0}\right)^{\frac{1}{2}k_1k_3} \\
\nonumber
&&
\times \int d^2u \left|\frac {du}{dx}\right|^{-2}
\left|\frac{x\8}{\9}\right|^{\frac{1}{2}k_1k_4}
\left|\frac{(x-1)\7}{\9}\right|^{\frac{1}{2}k_3k_4}
T_{IJKL}^{i_1i_2i_3i_4}(u)T_{IJKL}^{j_1j_2j_3j_4}(\bar{u}),
\eea
where we have introduced a concise notation
$$
T_{IJKL}^{i_1i_2i_3i_4}(u)=\l\t_i\t_j\r T_{IJKL}^{i_1i_2i_3i_4 ij}(u).
$$
There is an important subtlety originating from the non-abelian 
nature of the orbifold model that leads to changing the 
overall normalization of (\ref{int}) by some constant $\kappa$.
Recall that our computation scheme relies on independent
computation of boson and fermion (holomorphic and antiholomorphic)
contributions to the correlation function (\ref{guu}), i.e. we
regard the vertex operators as the tensor product of bosonic 
and fermionic, and holomorphic, and antiholomorphic twist fields.
However, as was mentioned above, the absence of such a decomposition
in the orbifold model has to be taken into account.
A correct way of determining the normalization constant for the 
correlation functions should be based on the OPE (\ref{OPE}) of the vertex operators
and involves the knowledge of the corresponding structure constants.
Namely, omitting all unessential details,
the normalization constant turns out to be proportional
to the product of two structure constants: 
$C(g_0,g_\infty)\sim \check{C} \check{C}$, while $\check{C}$  
are obtained by considering an auxiliary correlation function 
$\l V_{g_0^{-1}}(\infty)V_{IJ}(1)V_{IJ}(u)V_{g_0}(0)\r $.  It's 
normalization is found by studying the limit $u\to 1$ and does not 
appeal to the tensor product structure of the vertex 
operators. In the limit $u\to 0$ one gets 
$\l V_{g_0^{-1}}(\infty)V_{IJ}(1)V_{IJ}(u)V_{g_0}(0)\r\to 
\frac{f}{u^{\D}\bar{u}^{\bar{\D}}}$, where $f$ is some constant,
which due to (\ref{OPE}) is related with $\check{C}$ as
$f=2 \check{C}^2$. Note that the multiplier 2 emerges namely  due to
the nonabelian character of the orbifold.
On the other hand, the constant $f$ is 
expressed through the structure constants $C_{bh},C_{fh}$ and 
$C_{ba},C_{fa}$ \footnote{Here, e.g., $C_{bh}$ refers to 
the structure constant (\ref{strcons}) in the bosonic
left-moving sector.} found in 
the previous sections as $f=2^4C_{bh}^2C_{fh}^2C_{ba}^2C_{fa}^2$, 
since each sector again provides the multiplier $2$. Thus, 
the structure constant $\check{C}$
is equal to the product $\check{C}=2^{3/2}C_{bh}C_{fh}C_{ba}C_{fa}$.
Therefore, the constant $\kappa$ is found to be $2^3$.

Coming back to eq.(\ref{int}), we
see that under the change of variables $u\to x$
the integral acquires the form
\bea
\nonumber
&&\int d^2u |u| G_{IJKL}^{i_1i_2i_3i_4}(u)\bar{G}_{IJKL}^{j_1j_2j_3j_4}(\bar{u})
=\frac{R^8}{2^{9}}\frac{n_0n_\infty(N-n_\infty)}{(N-n_0)^3}
\left(\frac{n_\infty-n_0}{N-n_0}\right)^{\frac{1}{2}k_1k_3} \\
\nonumber
&&\times
\int _{V_{IJKL}} d^2x  
\left|\frac{x\8}{\9}\right|^{\frac{1}{2}k_1k_4}
\left|\frac{(x-1)\7}{\9}\right|^{\frac{1}{2}k_3k_4}
T^{i_1i_2i_3i_4}(x)T^{j_1j_2j_3j_4}(\bar{x}),
\eea
where we have taken into account that under this change of variables the
$u$-sphere is mapped onto the domain $V_{IJKL}$.

Since the basic
correlation function (\ref{fuu1}) is equal to the sum
$$
F(u,\bar u)=
\frac {C_0C_\infty }{N!}2n_0(N-n_0)n_\infty (N-n_\infty )
\sum_{IJKL}
G_{IJKL}^{i_1i_2i_3i_4}(u)\bar{G}_{IJKL}^{j_1j_2j_3j_4}(\bar{u})
\zeta_1^{i_1j_1}\zeta_2^{i_2j_2}\zeta_3^{i_3j_3}\zeta_4^{i_4j_4},
$$
where the summation goes over the set of indices listed in eq.(\ref{s2}),
the integral $\int d^2u|u|F(u,\bar u)$ is equal to
\bea
\la{int2}
&&\int d^2u|u|F(u,\bar u)=
\frac{R^8}{2^{8}\sqrt{k_1^+k_2^+k_3^+k_4^+}}
\left(\frac{n_0n_\infty(N-n_\infty)}{N(N-n_0)}\right)^2
\left(\frac{n_\infty-n_0}{N-n_0}\right)^{\frac{1}{2}k_1k_3} \\
\nonumber
&&\times
\int d^2x  
\left|\frac{x\8}{\9}\right|^{\frac{1}{2}k_1k_4}
\left|\frac{(x-1)\7}{\9}\right|^{\frac{1}{2}k_3k_4}
T^{i_1i_2i_3i_4}(x)T^{j_1j_2j_3j_4}(\bar{x})
\zeta_1^{i_1j_1}\zeta_2^{i_2j_2}\zeta_3^{i_3j_3}\zeta_4^{i_4j_4}.
\eea

To discuss the Lorentz invariance of the theory, without loss of generality,
we choose the polarization $\z^{\mu\nu}$ in the form $\z^{\mu}\z^{\nu}$.
Recall that in ten dimensions a polarization of a graviton satisfies
the transversality condition: $k_{\mu}\z^{\mu}=0$. In the light-cone
gauge the polarization obeys $\z^{+}=0$ allowing to express 
the component $\z^{-}$ via $\z^{i}$ and $k_{\mu}$ as 
$\z^{-}=\frac{2k^i\z^i}{k^+}$. In our model we only deal with eight
transversal polarizations $\z^i$ and can treat the last equation
as a definition of the light-cone polarization $\z^-$. An important
property of the light-cone gauge is that 
$\z_1^{i}\z_2^{i}=\z_1^{\mu}\z_2^{\mu}\equiv (\z_1\z_2)$.
Glearly, the integrand in (\ref{int2}) depends on scalar products
of the transversal momenta $k^i$ and polarizations $\z^i$.
It turns out that by using the light-cone momenta and polarizations
$k^-$ and $\z^-$ the integrand can be written via 
scalar products of ten-dimensional vectors.
Namely, it is enough to check this assertion for the expressions
$\l \t_i \t_i\r$ and $\l \t_i \t_j\r \z_a^i \z_b^j$. We confine
ourselves with considering $\l \t_i \t_j\r \z_a^i \z_b^j$.
The result easily follows from the fact that, e.g.,
\bea
\nonumber
&&\frac{\7}{n_0}k_1^i \z_a^i+\frac{\8}{n_\infty} k_3^i\z_a^i
+\frac{1}{N-n_0}k_4^i\z_a^i \\
\nonumber
&&=
\frac{\7}{n_0}(k_1\z_a )+\frac{\8}{n_\infty} (k_3\z_a)
+\frac{1}{N-n_0}(k_4\z_a),
\eea
where the relation 
$k^i\z^i_a=(k\z_a)+\frac{1}{2}k^+\z_a^-$ was used.

To rewrite the integral (\ref{int2}) in the conventional form 
we perform the change of variables
\bea
\nonumber
\frac {n_\infty -n_0}{N-n_0} z=\frac 
{x(x-\frac {N-n_0-n_\infty }{N-n_0})}{x-\frac {n_0}{n_0-n_\infty }},
\qquad
dz=\frac{(x-\al_1)(x-\al_2)}{\frac {n_\infty -n_0}{N-n_0}\9 ^2}.
\eea
Then, after simple but rather lengthy calculations, one arrives at
the following result
\bea
\nonumber
\int d^2u|u|F(u,\bar u)=
\frac{R^8}{2^{8}N^2\sqrt{k_1^+k_2^+k_3^+k_4^+}}
\int d^2 z  
\left|z\right|^{\frac{1}{2}k_1k_4-2}
\left|1-z\right|^{\frac{1}{2}k_3k_4-2}
K(z,\bar{z},\z),
\eea
Here we introduced a notation
$$
K(z,\bar{z},\z)=
K^{i_1i_2i_3i_4}(z)K^{j_1j_2j_3j_4}(\bar{z})
\zeta_1^{i_1j_1}\zeta_2^{i_2j_2}\zeta_3^{i_3j_3}\zeta_4^{i_4j_4},
$$
where 
\bea
\nonumber
&&K^{i_1i_2i_3i_4}(z)\zeta_1^{i_1}\zeta_2^{i_2}\zeta_3^{i_3}\zeta_4^{i_4}=\\
\nonumber
&& \qquad
z(k_3k_4)(\z_1\z_3)(\z_2\z_4)
-(k_3k_4)(\z_2\z_3)(\z_1\z_4)
-(k_1k_4)(\z_1\z_2)(\z_3\z_4) \\
\nonumber
&& \qquad +
(1-z)\left[
(\z_1k_4)(\z_3 k_2)(\z_2\z_4)+
(\z_2k_3)(\z_4 k_1)(\z_1\z_3)+
(\z_1k_3)(\z_4 k_2)(\z_2\z_3)+
(\z_2k_4)(\z_3 k_1)(\z_1\z_4) \right]\\
\nonumber
&&\qquad +z\left[
(\z_2k_1)(\z_4 k_3)(\z_3\z_1)+
(\z_3k_4)(\z_1 k_2)(\z_2\z_4)+
(\z_2k_4)(\z_1 k_3)(\z_3\z_4)+
(\z_3k_1)(\z_4 k_2)(\z_1\z_2)
\right] \\
\nonumber
&&\qquad -\left[
(\z_1k_2)(\z_4 k_3)(\z_3\z_2)+
(\z_3k_4)(\z_2 k_1)(\z_1\z_4)+
(\z_1k_4)(\z_2 k_3)(\z_3\z_4)+
(\z_3k_2)(\z_4 k_1)(\z_1\z_2)
\right]. 
\eea
Now one can recognize in $K$ the standard open string kinematical 
factor for the four vector particle scattering.

The S-matrix element can be now found by using eq.(\ref{matel3})  
and by taking the limit $R\to\infty$:
\bea
\langle f|S|i\rangle =
-i\frac{\lambda^2 N\d_{m_1+m_2+m_3+m_4,0}\d (\sum_i k_i^-)\d^D(\sum_i \k_i)}
{2^7\sqrt {k_1^+k_2^+k_3^+k_4^+}}
\int d^2z 
\left|z\right|^{\frac{1}{2}k_1k_4-2}
\left|1-z\right|^{\frac{1}{2}k_3k_4-2}K(z,\bar{z},\z),
\label{77}
\eea
where we have restored $\d$-functions responsible for the momentum 
conservation law and have represented
the light-cone momenta $k_i^+$ as $k_i^+=\frac {m_i}{N}$.

In the limit $N\to\infty$ the combination $N\d_{m_1+m_2+m_3+m_4,0}$ goes to 
$\d (\sum_i k_i^+)$ and eq.(\ref{77}) acquires the form
\bea
\langle f|S|i\rangle =-i\frac{\lambda^2
\d^{D+2}(\sum_i k_i^\mu )}
{2^{8}\sqrt {k_1^+k_2^+k_3^+k_4^+}}
\int d^2z 
\left|z\right|^{\frac{1}{2}k_1k_4-2}
\left|1-z\right|^{\frac{1}{2}k_3k_4-2}K(z,\bar{z},\z).
\nonumber
\eea
Taking into account that the scattering amplitude $A$ is related to the
S-matrix element as follows (see e.g. \cite{GSW})
$$
\langle f|S|i\rangle =-i\frac{\d^{D+2}(\sum_i k_i^\mu )}
{\sqrt {k_1^+k_2^+k_3^+k_4^+}}A(1,2,3,4),$$
one finally gets
$$
A(1,2,3,4)=\lambda^22^{-8}\int d^2z
\left|z\right|^{\frac{1}{2}k_1k_4-2}
\left|1-z\right|^{\frac{1}{2}k_3k_4-2}K(z,\bar{z},\z)
$$
that is the well-known four graviton scattering amplitude.

\section{Conclusion}
In this paper we developed the technique for calculating scattering 
amplitudes of type II string states by using the interacting 
$S^N\R^{8}$ orbifold sigma model. Although we considered only the 
four graviton scattering, our results allow one to find immediately
the scattering amplitudes of any four particles.
Let us stress that in our calculation
we did not impose any kinematical restrictions
on momenta and polarizations of gravitons and, hence, 
automatically obtained the Lorentz-invariant amplitude. 
It gives a strong evidence that the two-dimensional 
Yang-Mills model should possess the same invariance in the large $N$ limit.

An interesting problem is to consider the 
scattering amplitudes of the heterotic string states.
Recall that an important point of our construction was
the cancellation of phases coming both from the left- and 
the right-moving sectors. The world-sheet parity transformations
combined with an odd number of space reflections are the symmetry
of the type IIA string theory responsible for this cancellation.
A symmetry of the heterotic string theory that may be responsible 
for such a cancellation is unknown.

It would be of interest to trace the appearance
of the loop amplitudes in the framework of the orbifold 
sigma model. Obviously, the one-loop amplitude requires the 
computation of the correlation function of four DVV interaction vertices
sandwiched between the asymptotic states, which
technically results in constructing
the non-commutative Green functions in the presence of six twist
fields. We note that cancellation of possible divergences in the 
amplitude may require the further perturbation of the CFT action by 
higher-order contact terms.

Although the string scattering amplitudes follow from the orbifold
model description, the problem of a great interest is to reproduce
the amplitudes directly from the SYM theory.

{\bf ACKNOWLEDGMENT} The authors thank I.Y.Aref'eva, L.O.Chekhov, 
A.Yu.Morozov and A.A.Slavnov for valuable discussions. 
This work has been supported in part 
by the RFBI grants N96-01-00608, N96-01-00551.  

\setcounter{section}{0}
\appendix{}
\setcounter{equation}{0}
We use the following representation of $\g$-matrices 
satisfying the relation 
$$
\g^i (\g^j)^T+\g^j (\g^i)^T=2\d^{ij}I
$$
\bea
\nonumber
\begin{array}{ll}
\gamma^1=
\left(\begin{array}{rr}1&0\\0&1\end{array}\right)\otimes
\left(\begin{array}{rr}0&1\\1&0\end{array}\right)\otimes
\left(\begin{array}{rr}0&1\\-1&0\end{array}\right)  &
\gamma^2=
\left(\begin{array}{rr}0&1\\-1&0\end{array}\right)\otimes
\left(\begin{array}{rr}0&1\\-1&0\end{array}\right)\otimes
\left(\begin{array}{rr}0&1\\-1&0\end{array}\right)  \\
\gamma^3=1 &
\gamma^4=
\left(\begin{array}{rr}0&1\\-1&0\end{array}\right)\otimes
\left(\begin{array}{rr}1&0\\0&1\end{array}\right)\otimes
\left(\begin{array}{rr}1&0\\0&-1\end{array}\right) \\
\gamma^5=
\left(\begin{array}{rr}1&0\\0&1\end{array}\right)\otimes
\left(\begin{array}{rr}1&0\\0&-1\end{array}\right)\otimes
\left(\begin{array}{rr}0&1\\-1&0\end{array}\right)
& 
\gamma^6=-
\left(\begin{array}{rr}0&1\\-1&0\end{array}\right)\otimes
\left(\begin{array}{rr}1&0\\0&1\end{array}\right)\otimes
\left(\begin{array}{rr}0&1\\1&0\end{array}\right) \\
\gamma^7=
\left(\begin{array}{rr}1&0\\0&-1\end{array}\right)\otimes
\left(\begin{array}{rr}0&1\\-1&0\end{array}\right)\otimes
\left(\begin{array}{rr}1&0\\0&1\end{array}\right) 
& 
\gamma^8=
\left(\begin{array}{rr}0&1\\1&0\end{array}\right)\otimes
\left(\begin{array}{rr}0&1\\-1&0\end{array}\right)\otimes
\left(\begin{array}{rr}1&0\\0&1\end{array}\right).
\end{array}
\eea
Consider the spinor representation ${\bf 8_s}$ of the $SO(8)$.
In this representation the algebra generators $R^{ij}$ can be realized as
$$
R^{ij}=\frac{1}{4}\theta^{a}\gamma^{ij}_{ab}\theta^{b}.
$$
Here $\gamma^{ij}=\frac{1}{2}(\gamma^i(\gamma^j)^T-\gamma^j(\gamma^i)^T)$.

According to our definition of the fermions $\Theta$:
$$
\Theta^A=\frac{\theta^{A}+i\theta^{A+4}}{\sqrt{2}},~~~
\Theta^{\bar{A}}=\frac{\theta^{A}-i\theta^{A+4}}{\sqrt{2}},~~~
$$
where $A=1,\ldots,4$, we get for $R^{ij}$:
$$
R^{ij}=\frac{1}{4}(\Theta^AS^{ij}_{AB}\Theta^B+
\Theta^{\bar{A}}\bar{S}^{ij}_{AB}\Theta^{\bar{B}}+
\Theta^{\bar{A}}\bar{T}^{ij}_{AB}\Theta^B+
\Theta^A T^{ij}_{AB}\Theta^{\bar{B}},
$$
where
$$
T^{ij}_{AB}=\frac{1}{2}
(\gamma^{ij}_{AB}+i\gamma^{ij}_{A~B+4}
-i\gamma^{ij}_{A+4~B}+\gamma^{ij}_{A+4~B+4})
$$
and
$$
S^{ij}_{AB}=\frac{1}{2}
(\gamma^{ij}_{AB}-i\gamma^{ij}_{A~B+4}
-i\gamma^{ij}_{A+4~B}-\gamma^{ij}_{A+4~B+4}).
$$

The four Cartan generators of $SO(8)$ are given by $R^{2i-1~2i}$. 
Define the $SU(4)\times U(1)$ generators acting on $\Theta^A$s':
$$
J^{ij}=-\frac{i}{2}\Theta^A T^{ij}_{AB}\Theta^{\bar{B}}.
$$
Then the Cartan generators $H_1,H_2,H_3$ of $SU(4)$ 
and the generator $H_4$ of $U(1)$ are given by
$$
H^1= J^{12},~~H^2= J^{34},~~H^3= J^{56},~~H^4 = J^{78}.
$$
Since 
$$
T^{12}=i
\left(\begin{array}{cccc}
 -1&  &  &    \\
  & -1&  &    \\
  &  & 1&    \\
  &  &  &1
\end{array}\right)~~~~
T^{34}=i
\left(\begin{array}{cccc}
 -1&  &  &    \\
  & 1&  &    \\
  &  & -1&    \\
  &  &  &1
\end{array}\right)~~~~
T^{56}=i
\left(\begin{array}{cccc}
1&  &  &    \\
  & -1&  &    \\
  &  & -1&    \\
  &  &  &1
\end{array}\right)~~~~
T^{78}=i I
$$
the weights of the  representation ${\bf 4_{1/2}}$ look as
\bea
\nonumber
&&q^1=\frac{1}{2}(-1,-1,1,1);~~~q^3=\frac{1}{2}(1,-1,-1,1);\\
&&q^2=\frac{1}{2}(-1,1,-1,-1);~~~q^4=\frac{1}{2}(1,1,1,1);
\eea
Bosonizing the Cartan generators with the help of four bosonic fields $\phi^A$
as $H^A=i\partial \phi^A$ we get the following expression for fermions
$$
\Theta^A=e^{iq_B^A\phi^B}.
$$

\appendix{}
\setcounter{equation}{0}
In this Appendix we consider some properties of the map (\ref{map}) and outline
the derivation of the differential equation (\ref{difur4}) for the four-point
correlation functions (\ref{guu}). 

Let us consider the map (\ref{map}) 
\be 
z=\frac {t^{n_0}(t-t_0)^{N-n_0}}{(t-t_\infty )^{N-n_\infty}} 
\frac {(t_1-t_\infty )^{N-n_\infty}}{t_1^{n_0}(t_1-t_0)^{N-n_0}} 
\equiv u(t).  
\la{a1} 
\ee 
This map is the $N$-fold covering of the $z$-sphere by the 
$t$-sphere. Obviously, it branches at the points $t=0,t_0,t_\infty$ 
and $\infty$. To find other branch 
points we have to solve the following equation:  \bea \frac{d\log 
z}{dt}&=& 
\frac{n_0}{t}+\frac{N-n_0}{t-t_0}-\frac{N-n_\infty}{t-t_\infty}\nonumber\\
&=&\frac{n_\infty t^2+\left( (N-n_0-n_\infty )t_0-Nt_\infty \right) t+
n_0t_0t_\infty}{t(t-t_0)(t-t_\infty )}.
\la{a2}
\eea
In general there are two different solutions $t_1$ and $t_2$ of this equation,
and the map (\ref{a1}) has the following form in the vicinity of these points
\bea
z-z_i\sim (t-t_i)^2 ,\quad z_1=1=u(t_1),\quad z_2=u=u(t_2).
\nonumber
\eea
Due to the projective transformations, we can impose three relations on
positions of branch points. However, we have already chosen the points $0$ and
$\infty$ as two branch points, therefore, only one relation remains to be
imposed. Since the differential equation on the four-point correlation function is
written with respect to the point $u$, it is convinient not to fix the position
of the point $t_2\equiv x$. Then, the remaining relation that leads to the
rational dependence of points $t_0,t_\infty$ and $t_1$ on $x$ looks as follows
\be
t_0=x-1.
\la{a4}
\ee
The point $x$ is supposed to be a solution of eq.(\ref{a2}). Therefore, one can
immediately derive from eqs.(\ref{a2}) and (\ref{a4}) that  $t_\infty$ is 
expressed through the point $x$ as
\be
t_\infty =x-\frac {(N-n_\infty )x}{(N-n_0)x+n_0}.
\la{a5}
\ee
The second solution of eq.(\ref{a2}) can be now easily found and is given by
\bea
t_1&=&\frac {N-n_0-n_\infty }{n_\infty}+\frac {n_0x}{n_\infty}-
\frac {N(N-n_\infty )x}{n_\infty ((N-n_0)x+n_0)}\nonumber\\
&=&\frac{n_0(x-1)\left( (N-n_0)x+n_0+n_\infty -N\right)}
{n_\infty\left( (N-n_0)x+n_0\right)}.
\la{a6}
\eea
The rational function $u(x)$ is defined by the following equation
\be
u(x)=\frac {x^{n_0}(x-t_0)^{N-n_0}(t_1-t_\infty )^{N-n_\infty}}
{(x-t_\infty )^{N-n_\infty}t_1^{n_0}(t_1-t_0)^{N-n_0}}.
\la{a7}
\ee
By using eqs.(\ref{a4}),(\ref{a5}) and (\ref{a6}), one can derive the following
relations
\bea
t_1-t_0&=&\frac{(N-n_0)(x-1)\left( (n_0-n_\infty )x-n_0\right)}
{n_\infty\left( (N-n_0)x+n_0\right)},\nonumber\\
t_1-t_\infty &=&\frac{\left( (n_0-n_\infty )x-n_0\right)
\left( (N-n_0)x+n_0+n_\infty -N\right)}
{n_\infty\left( (N-n_0)x+n_0\right)}.\nonumber
\eea
Then the rational function $u(x)$ is found to be equal to
\bea
u=u(x)&=&(n_0-n_{\infty})^{n_0-n_{\infty}}\frac{n_{\infty}^{n_{\infty}}}
{n_0^{n_0}}\left(\frac{N-n_0}{N-n_{\infty}}\right)^{N-n_{\infty}}
\left(\frac{x+\frac{n_0}{N-n_0}}{x-1}\right)^N\nonumber\\
&\times&\left(\frac{x-\frac{N-n_0-n_{\infty}}{N-n_0}}{x}\right)^{N-n_0-n_{\infty}}
\left(x-\frac{n_0}{n_0-n_{\infty}}\right)^{n_0-n_{\infty}}.
\la{a10}
\eea
To obtain the differential equation (\ref{difur4}) we need to know the
decomposition of the roots $t_K(z)$ and $t_L(z)$ in the vicinity of $z=u$. Let
us take the logarithm of the both sides of eq.(\ref{a1}):
\be
\log {\frac zu}=n_0\log {\frac tx}+(N-n_0)\log {\frac {t-t_0}{x-t_0}}-
(N-n_\infty )\log {\frac {t-t_\infty }{x-t_\infty }}.
\la{a11}
\ee
Decomposition of the l.h.s. of eq.(\ref{a11}) around $z=u$ and the r.h.s. 
of eq.(\ref{a11}) around $t=x$ gives:
\be
\sum_{k=1}^\infty \frac{(-1)^{k+1}}{k}\left(\frac{z-u}{u}\right)^k=
(t-x)^2\sum_{k=0}^\infty a_k(t-x)^k ,
\la{a12}
\ee
where the coefficients $a_k$ are equal to
\be
a_k=\frac {(-1)^{k-1}}{k+2}\left( \frac {n_0}{x^{k+2}}+
\frac {N-n_0}{(x-t_0)^{k+2}}-\frac {N-n_\infty}{(x-t_\infty)^{k+2}}\right) .
\la{a13}
\ee
It is clear from eq.(\ref{a12}) that $t(z)$ has the following decomposition
\be
t-x=\sum_{k=1}^\infty c_k(z-u)^{\frac k2}.
\la{a14}
\ee
Substituting eq.(\ref{a14}) into eq.(\ref{a12}), one finds
\bea
&&c_1^2=\frac {1}{ua_0},\quad c_2=-\frac {a_1}{2ua_0},\nonumber\\
&&2a_0c_1c_3=-\frac {1}{2u^2}+\frac {5a_1^2}{4u^2a_0^3}-
\frac {a_2}{u^2a_0^2}.
\la{a15}
\eea
Next coefficients are not important for us.

\noindent Then, by using the decomposition (\ref{a14}) and eq.(\ref{a15}), one
gets
\bea
&&\left(\frac {t''}{t'}\right) '=\frac {1}{2(z-u)^2}+O(1),\nonumber\\
&&\left(\frac {t''}{t'}\right)^2=\frac {1}{4(z-u)^2}+\frac {3}{z-u}
\left(\frac {c_2^2}{c_1^2}-\frac {c_3}{c_1}\right) +O(1),\nonumber\\
&&\frac {c_2^2}{c_1^2}-\frac {c_3}{c_1}=
\frac {1}{4u}\left( 1+\frac {2a_2}{a_0^{2}}-\frac {3a_1^2}{2a_0^3}\right) .
\nonumber
\eea
Finally, taking into account that in the set of $N$ roots $t_M(z)$ only two
roots $t_K(z)$ and $t_L(z)$ have the decomposition (\ref{a14}), we obtain
eqs.(\ref{dect}) and (\ref{difur3}).

The coefficients $a_k$ can be rewritten as the following functions of $x$:
\bea
a_0&=&\frac {n_0(n_0+n_\infty -N)}{2(N-n_\infty )x^2}+
\frac {n_0(N-n_0)}{(N-n_\infty )x}+
\frac {(N-n_0)(n_\infty -n_0)}{2(N-n_\infty )}\nonumber\\
&=&\frac {(N-n_0)(n_\infty -n_0)}{2(N-n_\infty )x^2}(x-\al_1)(x-\al_2),
\la{a18}\\
a_1&=&\frac {n_0((N-n_\infty )^2-n_0^2)}{3(N-n_\infty )^2x^3}-
\frac {n_0^2(N-n_0)}{(N-n_\infty )^2x^2}\nonumber\\&-&
\frac {n_0(N-n_0)^2}{(N-n_\infty )^2x}+
\frac {(N-n_0)((N-n_\infty )^2-(N-n_0)^2)}{3(N-n_\infty )^2},\nonumber\\
a_2&=&-\frac {n_0((N-n_\infty )^3-n_0^3)}{4(N-n_\infty )^3x^4}+
\frac {n_0^3(N-n_0)}{(N-n_\infty )^3x^3}+
\frac {3n_0^2(N-n_0)^2}{2(N-n_\infty )^3x^2}\nonumber\\
&+&\frac {n_0(N-n_0)^3}{(N-n_\infty )^3x}-
\frac {(N-n_0)((N-n_\infty )^3-(N-n_0)^3)}{4(N-n_\infty )^3}.\nonumber
\eea
To obtain the differential equation (\ref{difur4}) we have to use the following
important equalities on $\frac {1}{u}\frac {du}{dx}$, that can be derived by
using eqs.(\ref{a10}) and (\ref{a18})
\bea
\frac {1}{u}\frac {du}{dx}&=&\frac {n_0+n_\infty -N}{x}
-\frac {N}{x-1}+\frac {N}{x+\frac {n_0}{N-n_0}}\nonumber\\
&+&
\frac {N-n_0-n_\infty }{x-\frac {N-n_0-n_\infty }{N-n_0}}+
\frac {n_0-n_\infty }{x-\frac {n_0}{n_0-n_\infty }},\nonumber\\
\frac {1}{u}\frac {du}{dx}&=&\frac {4(N-n_\infty )^2x^4a_0^2}
{(N-n_0)^2(n_0-n_\infty )x(x-1)(x-\frac {N-n_0-n_\infty }{N-n_0})
(x-\frac {n_0}{n_0-n_\infty })(x+\frac {n_0}{N-n_0})}\nonumber\\
&=&\frac {(n_0-n_\infty )(x-\al_1)^2(x-\al_2)^2}
{x(x-1)(x-\frac {N-n_0-n_\infty }{N-n_0})
(x-\frac {n_0}{n_0-n_\infty })(x+\frac {n_0}{N-n_0})}.\nonumber
\eea
Finally, to get eq.(\ref{difur4}) one should use the Lagrange 
interpolation formula for the ratio of two polynomials 
\bea 
\frac{P(x)}{Q(x)}=\sum_i\frac{P(x_i)}{Q'(x_i)}\frac{1}{x-x_i},
\nonumber
\eea
where $x_i$ are the simple roots of $Q(x)$ and $degP<degQ$.

These equalities drastically simplify  the derivation of eq.(\ref{difur4}).


\end{document}